\documentclass[twocolumn]{aastex63}

\received{March 30, 2020}
\shorttitle{\ion{H}{i} clouds at high latitude}
\shortauthors{Panopoulou et al.}
\graphicspath{{./}{./}}

\usepackage{color}
\usepackage{url}
\usepackage[normalem]{ulem}

\usepackage{amsmath}	
\usepackage{amssymb}	

\DeclareMathAlphabet{\mathsc}{OT1}{cmr}{m}{sc}
\def\testbx{bx}%
\DeclareRobustCommand{\ion}[2]{%
\relax\ifmmode
\ifx\testbx\f@series
{\mathbf{#1\,\mathsc{#2}}}\else
{\mathrm{#1\,\mathsc{#2}}}\fi
\else\textup{#1\,{\mdseries\textsc{#2}}}%
\fi}

\begin{document}

\title{Maps of the number of \ion{H}{i} clouds along the line of sight at high galactic latitude}

\correspondingauthor{G. V. Panopoulou}
\email{panopg@caltech.edu}

\author[0000-0001-7482-5759]{G. V. Panopoulou}
\affiliation{California Institute of Technology, MC350-17, 1200 East California Boulevard, Pasadena, CA 91125, USA}
\altaffiliation{Hubble Fellow}

\author[0000-0001-5820-475X]{D. Lenz}
\affiliation{Jet Propulsion Laboratory, California Institute of Technology, Pasadena, California 91109, USA}



\begin{abstract}
Characterizing the structure of the Galactic Interstellar Medium (ISM) in three dimensions is of high importance for accurate modeling of dust emission as a foreground to the Cosmic Microwave Background (CMB). At high Galactic latitude, where the total dust content is low, accurate maps of the 3D structure of the ISM are lacking. We develop a method to quantify the complexity of the distribution of dust along the line of sight with the use of \ion{H}{i} line emission. The method relies on a Gaussian decomposition of the \ion{H}{i} spectra to disentangle the emission from overlapping components in velocity. We use this information to create maps of the number of clouds along the line of sight. We apply the method to: (a) the high-galactic latitude sky and (b) the region targeted by the BICEP/Keck experiment. In the North Galactic Cap we find on average 3 clouds per 0.2 square degree pixel, while in the South the number falls to 2.5. The statistics of the number of clouds are affected by Intermediate-Velocity Clouds (IVCs), primarily in the North. IVCs produce detectable features in the dust emission measured by \textit{Planck}. We investigate the complexity of \ion{H}{I} spectra in the BICEP/Keck region and find evidence for the existence of multiple components along the line of sight. The data (\url{https://doi.org/10.7910/DVN/8DA5LH}) and software are made publicly available, and can be used to inform CMB foreground modeling and 3D dust mapping.
\end{abstract}



\section{Introduction} \label{sec:intro}

The ability to reconstruct the 3D distribution of matter in the Galactic Interstellar Medium (ISM) is important for astrophysics and cosmology. 3D maps inform us about nearby Galactic spiral structure \citep[e.g.][and references therein]{rezaei2018}, dust evolution \citep[e.g.][]{schlaflyRv}, star forming regions \citep[e.g.][]{Zuckerdistances} and the ISM magnetic field \citep[e.g.][]{vaneck2017}. Such maps are also necessary for accurately modeling polarized dust emission \citep[e.g.][]{martinez-solaeche2018}, which acts as a dominant foreground contaminating the polarization of the Cosmic Microwave Background (CMB) \citep[e.g.][]{PlanckX2016}.

In recent years 3D mapping capabilities have improved drastically, largely due to the availability of large high-accuracy photometric datasets \citep[e.g. Pan-STARRS,][]{panstarrs} and information on stellar distances from \textit{Gaia} \citep{gaiamission}. Most 3D maps exploit the differential reddening of stars located at various distances to infer the 3D distribution of intervening ISM dust  \citep[e.g.][]{marshall2006,green2015, green2018, green2019,capitanio2017,lallement2014,lallement2019L,chen2019,leikeensslin2019}. A different, more classical method uses kinematic information from molecular (CO) and atomic (\ion{H}{i}) line emission as a tracer of the ISM density and relies on a model for the Galaxy's rotation curve to locate clouds as a function of distance \citep[e.g.][]{blitz1979,kulkarni1982,levine,Wenger2018}. It is possible to combine these two approaches, as demonstrated by \cite{kinetic_tomography2017}, to construct a 4D map of ISM clouds in the Galactic plane. 
These recent improvements in mapping complement our existing large-scale view of Galactic spiral arms \citep[obtained by astrometric measurements of masers in star forming regions, see][]{Reid2019} by revealing the Solar neighborhood structure in detail.

Existing 3D dust maps have complementary strengths \citep[for example higher angular resolution versus absence of fingers-of-god effect, see comparison in][]{green2019}. However, all share a common problem: they perform best at lower Galactic latitudes, whereas cosmology experiments avoid these high-dust-emission regions.
Maps that do cover regions far from the Galactic plane \citep[such as those by][]{lallement2019L,green2019} do not detect reddening towards many high-Galactic-latitude sightlines. This may arise partly from photometric uncertainties, which are comparable to the reddening in such regions \citep[e.g.][]{lenz2017}. The maps also do not extend to large distances at high Galactic latitude. The kinematic distance method does not apply in these parts of the sky, since cloud motions are affected by processes other than Galactic rotation \citep[e.g.][]{westmeier}.

This shortcoming is particularly troublesome for CMB science and the search for primordial B-mode polarization of the CMB \citep{kamionkowski,seljak}. Experiments that are searching for this signal target regions at high Galactic latitude where the Galactic foregrounds, such as dust and synchrotron emission, are lowest \citep[e.g.][]{BICEP2}. In order to separate the Galactic dust from the cosmological signal, this foreground emission must be modeled to great accuracy. An important challenge in this modeling comes from the fact that we do not know a priori the level of complexity of the ISM in these regions \citep[e.g. as discussed by][]{kiss}. The observed signal could arise from multiple dust components \citep{TassisPavlidou2015}, each with different properties (composition, temperature, magnetic field orientation). A large effort is underway to understand the effects of assumed dust models on the retrieval of cosmological parameters \citep[e.g.][]{remazeilles2016,Poh2017,PySM2017,Hensley2018}. Innovative, data-driven approaches are being developed to characterize the ISM properties \citep{clark2015,clark2018,philcox,gonzalezcasanova2019, guangyu}. Recent studies try to construct realistic realizations of dust foregrounds using 3D dust maps \citep{martinez-solaeche2018} or \ion{H}{i} data \citep{ghosh2017,adak2019,clarkhensley,Lu2019,hu2020}. 

When modeling dust in 3D, studies typically assume a model in which dust emission arises from discrete layers of dust along the line of sight \citep{plancklayers}. A key unknown in the modeling is the number of such layers that occupy a single pixel on the sky. Some studies assume an arbitrary value for the number of layers (clouds) along the line of sight \citep[e.g.][]{TassisPavlidou2015,Hensley2018}. Other studies adopt a more physically-motivated assumption about the number of clouds per pixel. For example, \cite{ghosh2017,adak2019} assume three dust layers, each arising from a different gas phase of the ISM. The approach of \cite{martinez-solaeche2018} assumes that each distance bin in the 3D dust map of \cite{green2019} is a different dust layer.

In this work we aim to \textit{measure} the number of clouds along each line of sight $-$ a critical but missing input to CMB foreground modeling. We focus on the high-Galactic latitude sky, where the 3D reconstruction of the dust distribution based on stellar reddening is incomplete. This difficulty can be surpassed by the use of an indirect, but more sensitive tracer of dust: \ion{H}{i} line emission. The \ion{H}{i} column density correlates well with dust in the diffuse ISM \citep[e.g.][]{boulanger1996, planckXI2014, lenz2017}. In contrast to existing 3D dust maps, \ion{H}{i} data show detection of the \ion{H}{i} line throughout the sky - there is no sightline free of \ion{H}{i} emission \citep{HI4PI}. 

Though the \ion{H}{i} line does not directly provide distances, important knowledge can be acquired through the kinematic information it carries. ISM clouds that lie at different distances likely have different kinematic properties, thus appearing as distinct kinematic components in \ion{H}{i} spectra.
This is especially true for the high Galactic latitude sky, where the \ion{H}{i} emission can be cleanly separated into three classes of clouds: Low, Intermediate and High - Velocity Clouds (LVC, IVC, HVC) based on their radial velocities with respect to the Local Standard of Rest ($\rm v_{LSR}$) \cite[e.g.][]{wakker1991,kunzdanly}.

Various approaches have been adopted to separate such kinematically distinct structures in \ion{H}{i} spectra.  \citet{haud2008} decompose \ion{H}{i} spectra into a set of Gaussian components using all-sky data from the Leiden/Argentine/Bonn survey \citep{kalberla2005}. They identify IVCs and HVCs by searching for overdensities in the parameter space of all Gaussian components found across the sky. Within smaller sky regions,  \citet{planck2011_xxiv} and \citet{martin2015} define the velocity range occupied by LVCs, IVCs, HVCs based on the standard deviation of the \ion{H}{i} spectrum. \citet{murray2019} 
construct smoothed \ion{H}{i} spectra towards the SMC by use of a Gaussian kernel and measure the number of peaks in the resulting spectra.

In this work we develop a method for cloud identification that builds on the complementary strengths of the aforementioned studies: it (a) locates overdensities in Gaussian component parameter space and (b) operates locally, within small sub-regions of the sky. The problem of identifying clouds by such means has been extensively addressed in the case of $\rm CO$ emission \citep[e.g.][]{clumpfind,dendrograms}. Recently, \cite{mamd2017} (hereafter MML17) developed a two-step method for cloud identification in $\rm CO$ spectral cubes. The MML17 method is based on a Gaussian decomposition of the data. It uses a hierarchical cluster analysis to identify clouds that are distinct in position-position-velocity (PPV) space. The MML17 method was applied to the \cite{dame2001} $\rm CO$ survey and resulted in a cloud catalog of unprecedented completeness.

The nature of the molecular medium allows for a well-defined process for cloud identification: $\rm CO$ is much more concentrated in space than \ion{H}{i} and allows one to draw boundaries to isolate distinct clouds \citep[e.g.][]{clumpfind}. Due to the much more diffuse nature of the atomic medium, a method that aims to define cloud boundaries, such as that of MML17, is not generally applicable to \ion{H}{i}. Exceptions include cases of very distant and/or compact clouds \citep[e.g.][]{Wakker1991III,Saul2012,Moss2013}, or restricting the analysis to the most cold and narrow \ion{H}{i} components, as done by \citet{haud2010}. However, we can use the basic principles of the MML17 approach for the specific problem of identifying distinct Low, Intermediate and High Velocity components far from the Galactic plane that lie along the \textit{same} line of sight. We develop a method that makes use of a Gaussian decomposition (as in many of the aforementioned works) but that only seeks to identify clouds within the same sightline and uses limited spatial information from neighboring pixels for this task.

We briefly describe the Gaussian decomposition in Section \ref{sec:decomp} and defer a detailed presentation in a separate paper \citep{lenz2019b}. The method developed here takes this decomposition as input and searches for prominent velocity components within groupings of neighboring pixels (as described in Section \ref{sec:method}). Preprocessing steps and validation are described in appendices \ref{sec:appendixA} and \ref{sec:validation}. We apply our method to regions of interest for CMB polarization studies: the sky at high Galactic latitude, and the region targeted by the BICEP/Keck experiment (Section \ref{sec:results}). We discuss our results in relation to CMB foreground analysis as well as the limitations of our method in Section \ref{sec:discussion}. The software and resulting data products are made publicly available and are presented in Section \ref{sec:product}. We summarize in Section \ref{sec:conclusion}.

\begin{figure}
\includegraphics[scale=1]{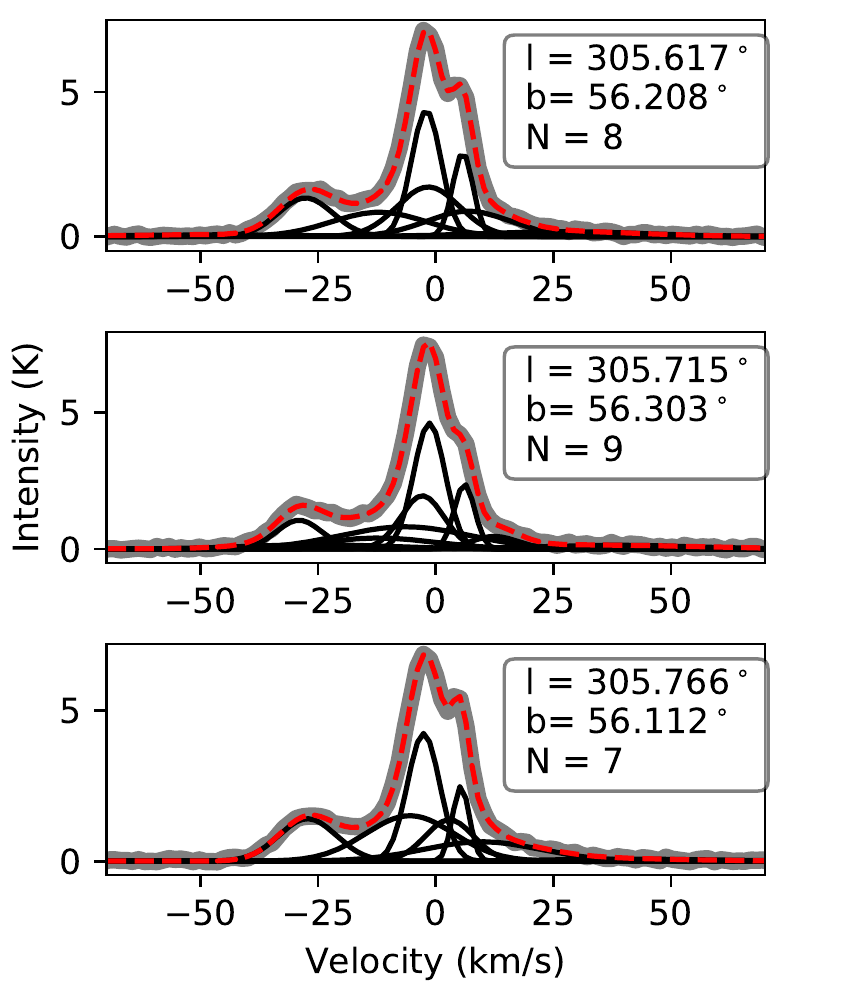}
\caption{Spectra of three neighboring pixels in the HI4PI dataset (gray lines) and Gaussian decomposition (each black line shows one Gaussian component). The sum of Gaussians in each spectrum is shown with a red dashed line. The box in the top right corner shows the Galactic coordinates of each pixel and the number of Gaussian components.}
\label{fig:spec}
\end{figure}

\begin{figure*}
\centering
\includegraphics[scale=1]{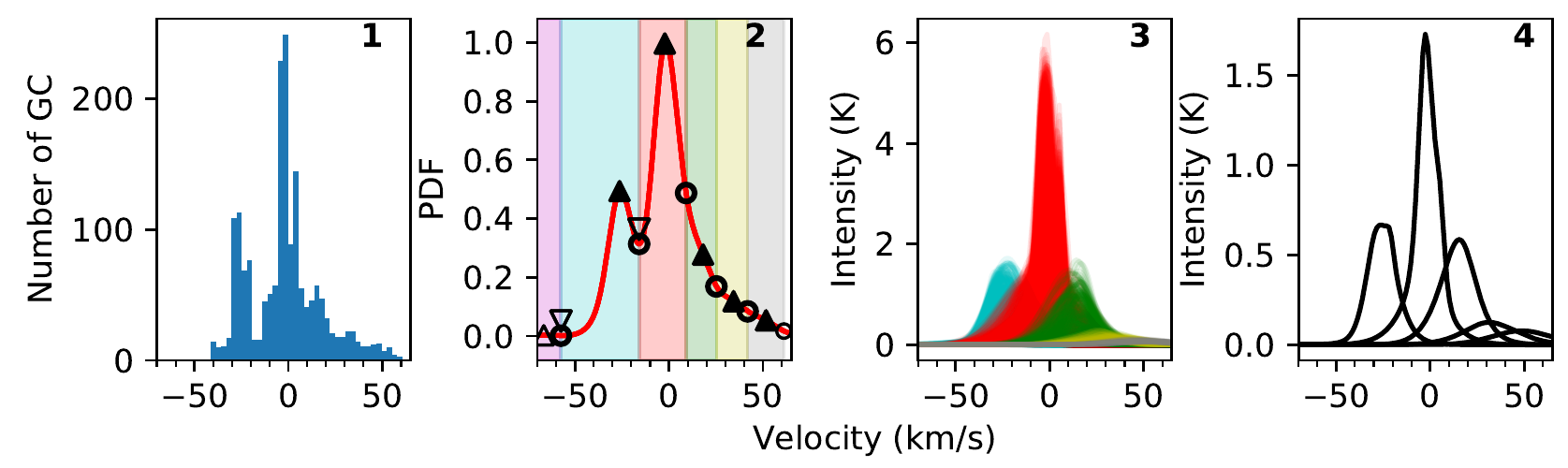}
\caption{\textit{Method of cloud identification}. Panel 1: Distribution of mean velocity of all Gaussian Components (GC) within a superpixel, after applying the quality cuts described in the text. Panel 2: KDE of the distribution in panel 1 (red line). Maxima (black upright triangles) and minima (open inverted triangles) of the PDF as well as maxima of the second derivative (open black circles) are used to define the location and velocity range of each peak. Peaks containing less than 20 GCs are later discarded (marked as empty upright triangles here). The velocity range of each peak is shown with a differently colored vertical band. Panel 3: Individual Gaussians colored according to the peak they belong to. Panel 4: Mean spectrum (average of GCs) for each peak.}
\label{fig:method}
\end{figure*}

\section{Data \& Gaussian decomposition}
\label{sec:decomp}

We use the all-sky survey of the \ion{H}{i} line presented in \cite{HI4PI} (HI4PI Survey). The dataset has an angular resolution of 16.2 arcmin and is sampled on a HEALPix grid of $N_{\rm{side}}$ 1024. It merges data from the Effelsberg-Bonn \ion{H}{i} Survey \citep[EBHIS,][]{winkel2010, kerp2011, winkel2016a} and the Galactic-All-Sky Survey \citep[GASS,][]{mcclure-griffiths2009, kalberla2010, kalberla2015} to create a full-sky database of Galactic atomic neutral hydrogen. The velocity range is  (-470, 470) km/s for the part of the sky covered by the Southern hemisphere survey, and (-600,600) km/s for the Northern hemisphere survey and the spectral resolution is 1.49 km/s (the channel separation is 1.29 km/s).

Each spectrum can be decomposed into a set of Gaussian basis functions, yielding a compressed description of the data \citep[e.g.][]{haud2000}. This approach has been adopted by many previous works for studying the properties of different ISM phases \citep[e.g.][]{haud2007,roy2013,gausspy,murray2017,kalberlahaud2018,marchal2019,gausspy+}, as well as detecting different classes of clouds \citep[e.g.][]{haud2008,haud2010}. 

\citet{lenz2019b} created a Gaussian decomposition of this dataset that is publicly available. For each spectrum, the intensity $I$ as a function of velocity $v$ is decomposed in a basis of the form: 
\begin{equation}
I({\rm v}) = \sum_i^N A_i \,\, \frac{1}{\sigma_i \sqrt{2\pi}}\,\, e^{\frac{({\rm v-v}_{0,i})^2}{2\sigma_i^2}},
\end{equation}
where $N$ is the number of Gaussian components (GCs), ${\rm v}_{0,i}$ is the centroid velocity of the $i$-th GC, $A_i$ is its amplitude, and $\sigma_i$ is its standard deviation. 
The spectrum is fit approximately 100 times with different numbers of components and initial parameters. The best compromise between complexity and goodness-of-fit is then selected via the Bayesian Information Criterion. The priors used for all three parameters of the GCs are uniform. $A$ is constrained so that the column density for each GC is within the range $[5\times 10^{18}\,, 5\times 10^{22}]\,\rm cm^{-2}$. The centroid velocity, ${\rm v}_{0}$, must lie within the HI4PI band. Finally, $\sigma$ is related to the \ion{H}{i} kinetic temperature, and is constrained to be within the range $[50\,, 4\times10^4]\,\rm K$. 
The model is fit to the velocity range [-300,+300] km/s and results in high-quality fits. We define the relative residual as the ratio of the difference between the model column density and the total column density, $\Delta N_{\ion{H}{i}}$, over the total column density of the pixel, $N_{\ion{H}{i}}$. The mean relative residual, is 1\%, while 65\% of pixels have a relative residual less than 5\%. 

Figure \ref{fig:spec} shows examples of the Gaussian decomposition for three neighboring pixels at high Galactic latitude. The original spectrum in each pixel is fit well by the sum of the Gaussian components. The three pixels show similar spectra, but have been fit by a different set of Gaussian components (the number of components varies from 7 to 9). This is to be expected, since Gaussians do not form an orthogonal basis.
The nonuniqueness of the fit gives rise to differences in the description of emission spectra that do not reflect actual changes in ISM properties \citep[e.g. as discussed by][]{marchal2019}. Such differences can be reduced by imposing spatial coherence criteria when performing the decomposition \citep[see][]{marchal2019}, with the disadvantage that this can be computationally expensive, especially for large sky areas. 
In the following section, we describe a method that overcomes this difficulty by combining the Gaussian decompositions from multiple neighboring pixels.

\begin{figure*}
\centering
\includegraphics[scale=0.9]{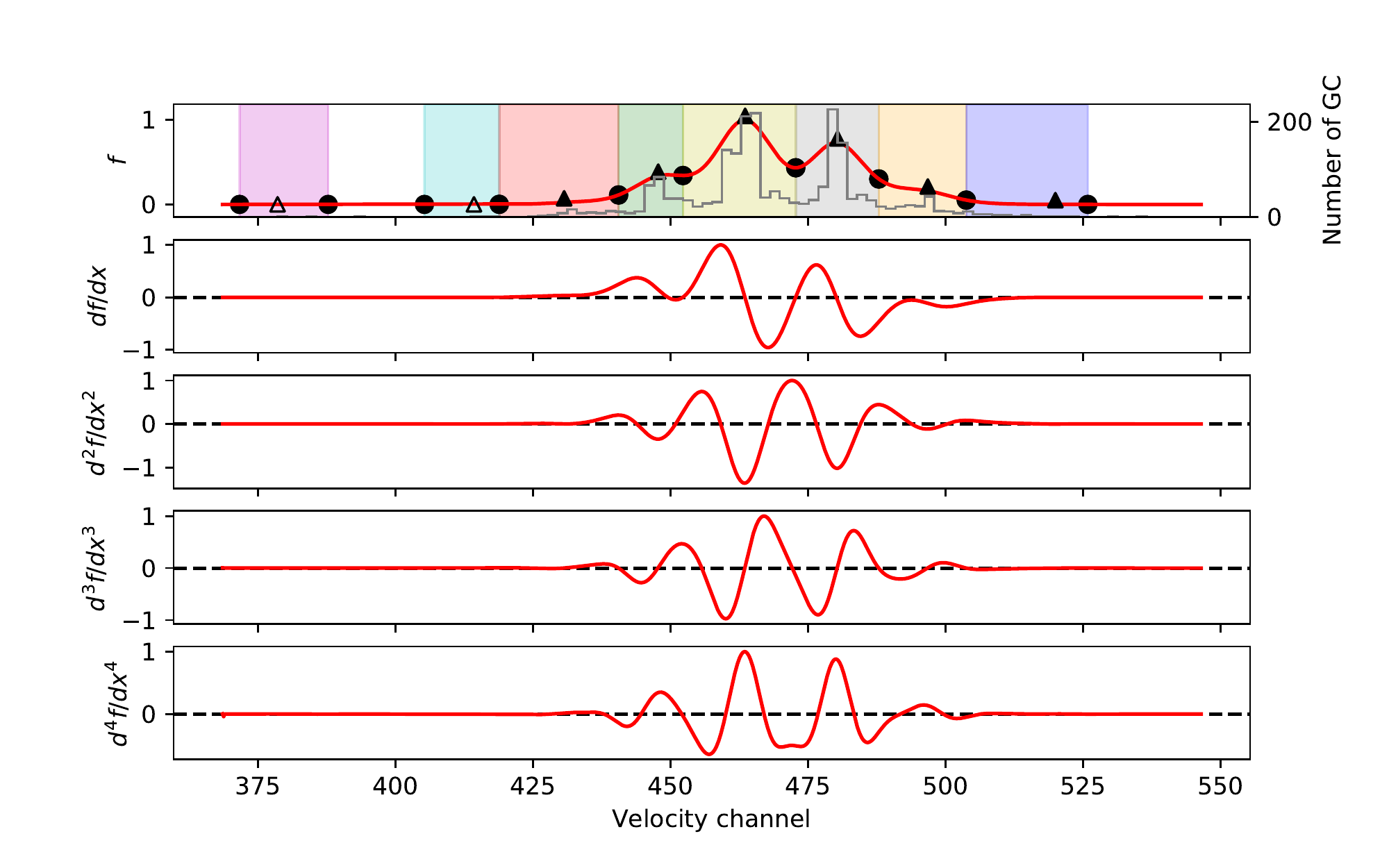}
\caption{\textit{Identifying peaks and corresponding velocity ranges in the PDF of Gaussian component velocities.} Top: Distribution of Gaussian component velocities in a superpixel (right axis, gray line) and constructed PDF using a KDE (left axis, red line). Maxima are marked with black upright triangles. Peaks with only a small number of GCs are discarded later in the analysis (marked here with empty upright triangles). The velocity range of each maximum is colored with a vertical band, its borders marked with black circles. Panels below show (in order) the 1st, 2nd, 3rd, 4th derivative of the PDF. The PDF and each of its derivatives are normalized to have a maximum equal to unity. Open upright triangles mark peaks with less than the threshold of 20 GCs. Peaks marked this way are spurious (only a small number of Gaussians are associated with them) and are removed via quality cuts.}
\label{fig:derivatives}
\end{figure*}

\section{Cloud identification}
\label{sec:method}


We define a cloud as a distinct peak appearing in the \ion{H}{i} spectrum. Peaks are often fit by multiple Gaussian components, the properties of which can vary between neighboring pixels (Figure \ref{fig:spec}). However, if Gaussians from multiple neighboring spectra appear to cluster around a certain region of parameter space (specifically in velocity space), then this is a strong indication that they are probing the same peak in the spectrum $-$ a cloud. This idea of searching for a `consensus' between adjacent pixels was used, for example, in the method of MML17 and in \citet{haud2010} to identify clouds in PPV space. Here, we adopt a different approach: we search for a `consensus' of GCs only in velocity space and consider GCs within a specified area, termed a superpixel. 

We collect information from multiple spectra by segmenting the sky into large superpixels, sampled on the HEALPix grid \citep{gorski2005,healpy}. The choice of superpixel size is a free parameter in the method. We have found that the method works well when there is a statistically significant number of Gaussian components in each superpixel. For illustrative purposes, we choose a superpixel size of  $N_{\rm{side}}$  = 128. Each superpixel thus contains 64 pixels of the HI4PI data. This superpixel size corresponds to 0.46$^\circ$ on each side, comparable to the angular resolution of the Lite satellite for the studies of B-mode polarization and Inflation from cosmic background Radiation Detection (LiteBIRD) at CMB frequencies \citep{litebird}. 
We investigate the effect of slightly varying the $N_{\rm{side}}$ parameter in Appendix \ref{sec:validation}.

We remove Gaussians associated with known artifacts. The part of the HI4PI survey that is based on the EBHIS data (Northern hemisphere) exhibits residual stray radiation patterns, which are inevitably fit by the Gaussian decomposition. We identify regions where this effect is more pronounced and remove the associated GCs as explained in Appendix \ref{sec:appendixA}. We also remove all GCs within the 13$\arcmin$ beam centered on (l, b) = (209.018$^\circ$, -19.37 $^\circ$) where the signal drops significantly to negative values.

Now that we have a set of Gaussian components that is free of artifacts, we proceed to analyse the distribution of mean velocities, $\nu_0$, within each superpixel (Figure \ref{fig:method}). We estimate the Probability Density Function (PDF) of $\nu_0$ using a Kernel Density Estimator (KDE) of Gaussian shape. The size of the kernel is selected through validation tests described in Appendix \ref{sec:validation}. The selected kernel standard deviation is 4 channels wide (5.1 km/s).

Next, we identify local maxima in the constructed PDF following the procedure described in \cite{gausspy}. We search for local minima of negative curvature (bumps), i.e. locations where all the following conditions are met:
\begin{itemize}
    \item The third derivative of the PDF changes sign.
    \item The second derivative is negative.
    \item The fourth derivative is positive.
\end{itemize}
Figure \ref{fig:derivatives} shows the PDF of $\nu_0$ and its derivatives up to fourth order in an example superpixel. Local maxima in the PDF indicate the presence of a `consensus': multiple Gaussians are tracing the same velocity component.

We now wish to assign each Gaussian to its corresponding peak. For this we must estimate the extent of each peak (the range of velocities belonging to each local maximum). To decide where a local maximum begins and ends we find the following three types of points:
\begin{itemize}
    \item[I.] Local minima of the PDF (sign change of the first derivative coincident with positive value of the second derivative),
    \item[II.] Points where the PDF is null (in practice, where its value is less than $10^{-4}$ of the maximum, meaning that there are no GCs at these velocities),
    \item[III.] Points where the second derivative is maximum. 
\end{itemize}
For each local maximum, we find the nearest point of any of the aforementioned types on either side of the maximum. We now have a velocity range assigned to each maximum (see Figure \ref{fig:method}, panel 2). Figure \ref{fig:method} shows an example PDF with these maxima and velocity ranges. In this example, the right border of the peak  at -30 km/s is a local minimum. The right border of the peak at -4 km/s is a point of type III, and coincides with the left border of the peak at +20 km/s. Type III points are useful when there is no minimum between two peaks. In cases where a point of type I or II is next to a point of type III with no maximum between them, points of the former types take precedence in border placement. This ensures there are no intervals where GCs exist but were not assigned to a peak.

It is important to note that not all peaks are good to be considered as clouds. First, when evaluating the PDF, edge effects can cause peaks at the beginning and end of the velocity range. We control for this by adding padding to the velocity axis prior to evaluating the PDF. Second, it is possible that a very small number of GCs is assigned to a given peak (for example the left-most peak in Figure \ref{fig:derivatives}). If there are less than 20 GCs that are associated with this peak, we discard the peak. We also discard clouds that cover less than 1 beam size (approximately 3 $N_{\rm{side}}$ 1024 pixels), which should remove point sources from our dataset.

We validate the code by testing it on mock data as presented in Appendix \ref{sec:validation}. We find that the method is able to recover clouds at the correct velocity for over 85\% of cases with a false positive rate of less than 10\%. We have also checked that our method performs well in assigning the majority of the \ion{H}{i} emission to clouds. Using the HI4PI data at high Galactic latitude, we demonstrate in Appendix \ref{sec:validation} that only a negligible number of pixels (1\%) show residual emission (not assigned to clouds) that is more than $10\%$ of the total column density. We note that our definition of a cloud is more general than that used by \cite{heilestroland} and includes not only CNM but also the WNM.

\begin{figure*}
\includegraphics[scale=1]{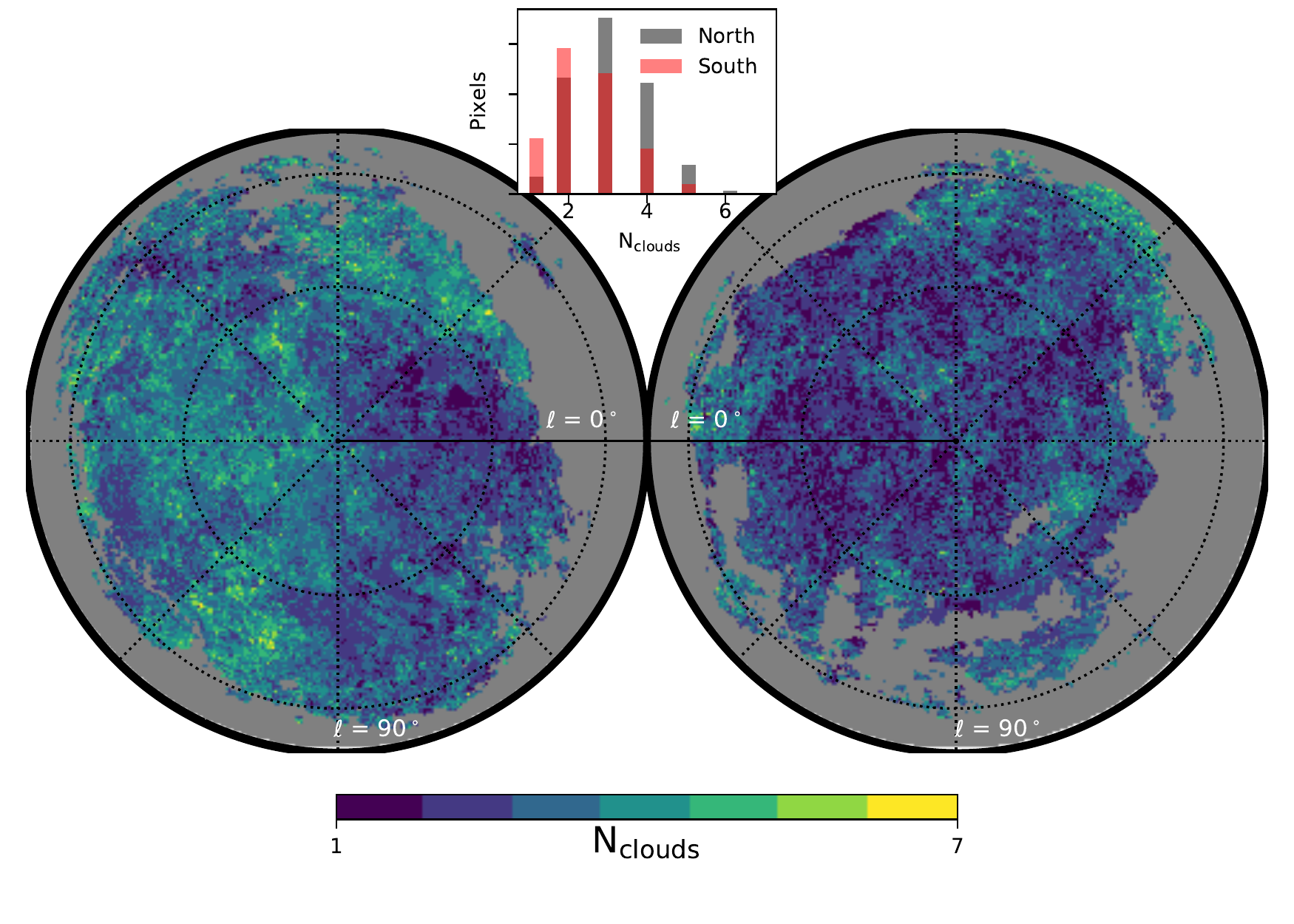}
\caption{Number of clouds ($\rm N_{clouds}$) per $N_{\rm{side}}$ 128 pixel (resolution $0.46$ deg) in the North (left) and South (right) regions in the velocity range $|\rm v_{LSR}| < 70 \rm{km/s}$. The maps are centered on the Galactic Poles and are in Orthographic projection. Parallels are spaced by 30$^\circ$ in latitude. The inset shows the 1D distribution of $\rm N_{clouds}$ per pixel for the North (gray) and South (red) regions.}
\label{fig:nclouds_map}
\end{figure*}

The method outputs the following information for each superpixel: (a) the number of identified clouds, (b) the parameters of all GCs that belong to each cloud. We post-process the output to compute cloud properties. For each cloud we calculate its mean column density in the superpixel, $N^{\rm Cloud}_{\ion{H}{i}}$, by averaging the column density of the cloud's GCs over all $N_{\rm{side}}$ 1024 pixels that they occupy. We construct a mean spectrum of the cloud, $<I({\rm v})>^{\rm Cloud}$, by averaging the intensities of all of the cloud's GCs at each velocity channel.
The centroid velocity of the cloud, $\rm v^{Cloud}$, is calculated as: 
\begin{equation}
    {\rm v}^{\rm{Cloud}}  = \frac{ \sum^{N_{\rm chan}}_i  {\rm v}_i \cdot <I({\rm v}_i)>^{\rm Cloud} }{\sum^{N_{\rm chan}}_i <I({\rm v}_i)>^{\rm Cloud} },
    \label{eqn:vcentroid}
\end{equation}
where ${\rm v}_i$ is the velocity of the $i$-th velocity channel and the summation is performed over all velocity channels ($N_{\rm chan}$) in the range [-300, 300] km/s. As a measure of the width of the cloud spectrum, we calculate the square root of its second moment:
\begin{equation}
{ \delta {\rm v}}^{\rm{Cloud}} = \sqrt{\sum^{N_{\rm chan}}_i \frac{<I({\rm v}_i)>^{\rm Cloud} \cdot ({\rm v}_i - {\rm v}^{Cloud})^2}{\sum^{N_{\rm chan}}_i <I({\rm v}_i)>^{\rm Cloud}} .}
\end{equation}

Since we retain the information on the GCs, we can also calculate quantities at the native resolution of the HI4PI data. For example, we compute the total $N_{\ion{H}{i}}$ per $N_{\rm{side}}$ 1024 pixel from GCs belonging to clouds. This is used to make morphological comparisons with higher resolution data (Section \ref{sec:results}) and to check the output of the algorithm, as explained in Appendix \ref{sec:validation}.

\section{Results}
\label{sec:results}

Our method is best suited for inferring 3D information about the dust distribution in regions with low dust content, where the \ion{H}{i} and dust column densities are linearly correlated \citep{planckXI2014}. Additionally, because the method relies on identifying kinematicaly distinct features, it is expected to perform best in areas where \ion{H}{i} spectra are relatively simple (with ideally 1 velocity component within the chosen KDE bandwidth).

For these reasons, we have chosen to apply our method to the high-Galactic latitude sky (Section \ref{ssec:Nclouds}), with column density $N_{\ion{H}{i}} < 4\times 10^{20} {\rm cm^{-2}}$, where \cite{lenz2017} find the best correspondence between \ion{H}{i} and dust emission. We also present results for the region targeted by the BICEP/Keck CMB experiment, as an illustration of the ability of our method to inform dust modeling for CMB foreground subtraction (Section \ref{sec:bicep}).

In our analysis we examine \ion{H}{i} emission within the velocity range $|\rm v_{LSR}| < 70 \rm{km/s}$, thus excluding HVCs which do not contribute traceable amounts of dust \citep[e.g.][]{wakker_boulanger,lenz2017}.

\begin{figure*}
\includegraphics[scale=1]{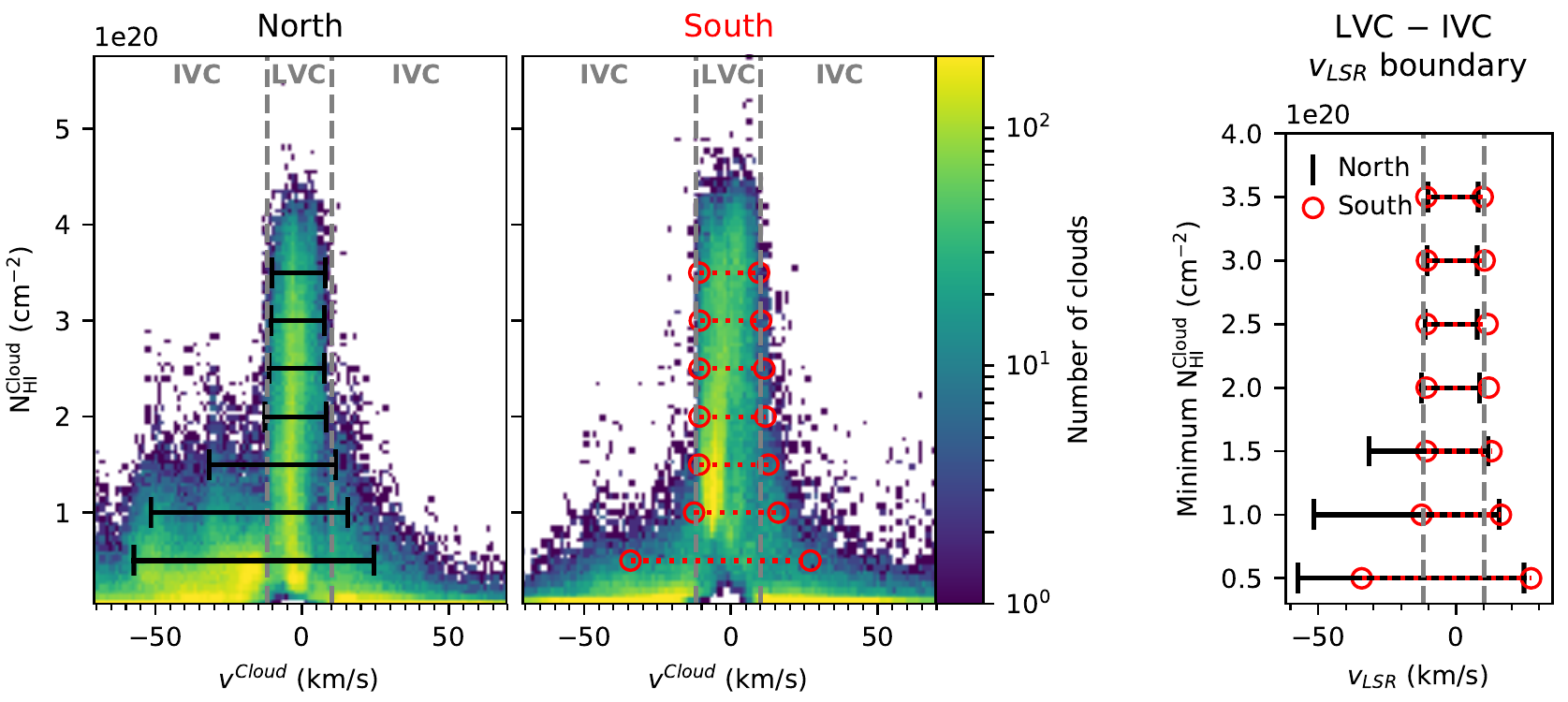}
\caption{\textit{Physical properties of identified clouds (Left two panels):}  Two-dimensional distribution of cloud \ion{H}{i} column density ($ N^{\rm Cloud}_{\ion{H}{i}}$) versus cloud centroid velocity for the North and South Galactic Polar regions.  The dashed vertical lines mark the velocity range that separates LVCs and IVCs (defined in the rightmost panel). Horizontal bars (solid black for the North and red dotted for the South) mark the 1$-$ and 99$-$percentile range of the distribution of centroid velocities of clouds above a minimum cloud $\rm N_{\ion{H}{i}}$ (see also right panel). \textit{Determination of velocity boundary that separates LVCs and IVCs (Right panel):}  Horizontal bars mark the 1$-$ and 99$-$ percentile range of cloud velocity for different thresholds of minimum cloud column density (black solid for clouds in the North, red dotted for clouds in the South). Dashed vertical gray lines mark the adopted $\rm v_{\rm LSR}$ boundaries at -12 and 10 km/s.}
\label{fig:nhvel}
\end{figure*}

\subsection{Polar areas}
\label{ssec:polarcaps}

We apply our method to all pixels in the HI4PI data with $N_{\ion{H}{i}} < 4\times 10^{20} {\rm cm^{-2}}$, which covers most of the sky at Galactic latitude $|b| > 30^\circ$. The total $N_{\ion{H}{i}}$ is calculated at the native resolution of the HI4PI data ($N_{\rm{side}}$ = 1024) and the resolution is then downgraded to produce a column density map at $N_{\rm{side}}$ = 128. Pixels of the low-resolution map that exceed the threshold in column density are masked. The total area covered is 17494 square degrees (42\% of the sky).

\subsubsection{The number of clouds per sightline at high Galactic latitude}
\label{ssec:Nclouds}

We produce maps of the number of clouds, $\rm N_{clouds}$, per $N_{\rm{side}} = 128$ pixel (resolution 0.46$^\circ$) (Figure \ref{fig:nclouds_map}). The maps are centered on the North and South Galactic Poles for better visualization. 
94\% of pixels have $\rm N_{clouds}$ = 1$-$4, with the maximum of 7 components found in a single pixel. There is no sightline free of clouds, as expected from the \ion{H}{i} spectra which show detections everywhere. The maps show large-scale coherence, despite the fact that each pixel has been treated independently. These large-scale regions of similar $\rm N_{clouds}$ per pixel, mark the presence of clouds that are spread out over hundreds of square degrees. The patterns are very different between the North and South Galactic Polar regions. 
We quantify this difference by looking at the 1D distribution of $\rm N_{clouds}$ per pixel for the North and South separately (inset of Figure \ref{fig:nclouds_map}, same data as in the maps). The distribution in the North has a mean of $\rm N_{clouds}$ = 3 compared to 2.5 in the South, and is skewed towards larger values.

\begin{figure*}
\includegraphics[scale=1]{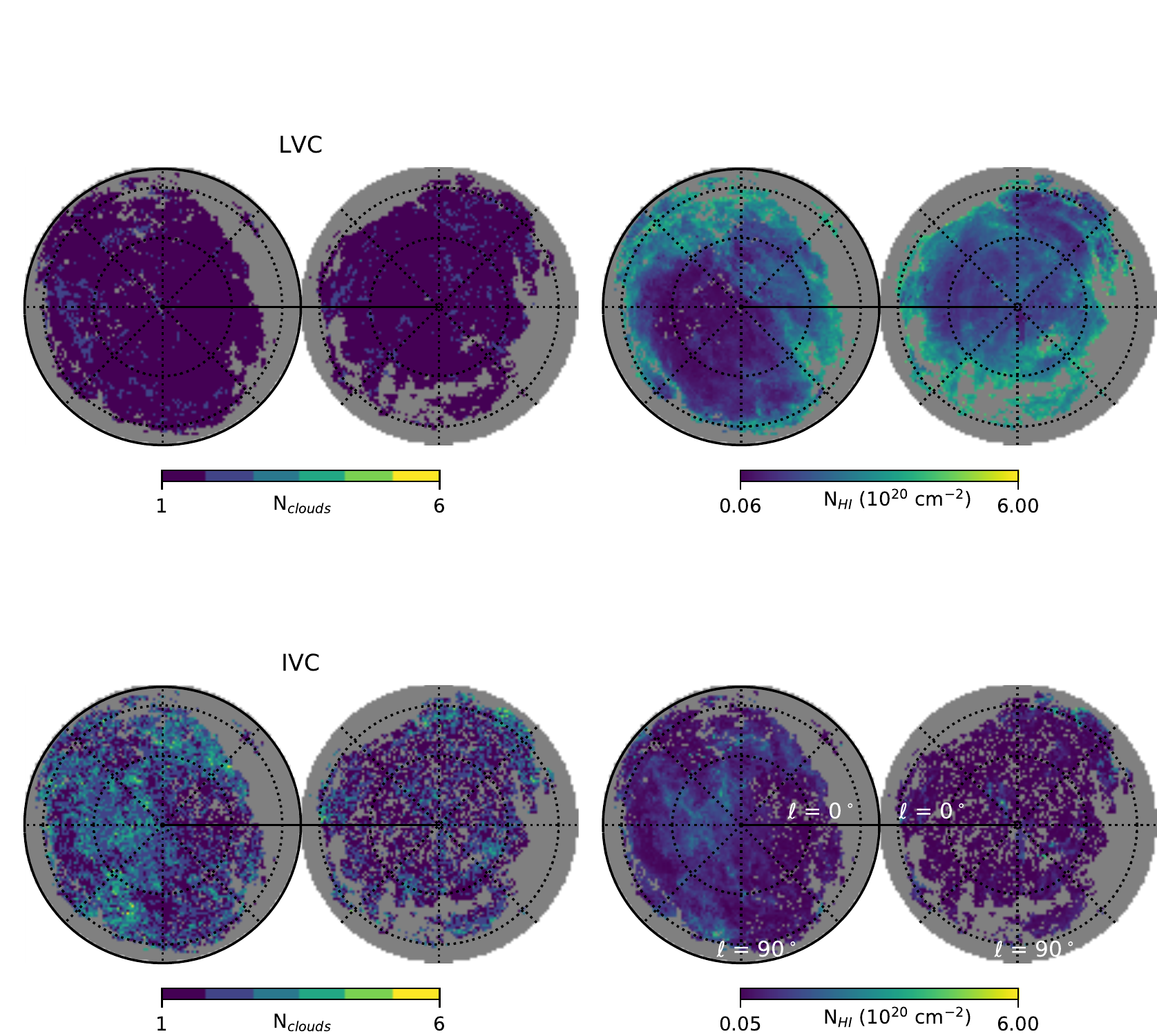}
\caption{Number of clouds per sightline (left panels) and column density (right panels) for clouds having a mean velocity in the LVC (top row) and IVC (bottom row) range. The maps are centered on the Galactic poles (North, left subpanel, and South, right subpanel), as in Figure \ref{fig:nclouds_map}. In all panels the colorscale is linear.}
\label{fig:NcloudsLVCIVC}
\end{figure*}

In order to understand these differences, we turn to the physical properties of the identified clouds. We examine the column density and centroid velocity of the clouds in Figure \ref{fig:nhvel} for the North and South regions separately. Clouds are found in the entire velocity range considered ($|\rm  v_{LSR}| <  70 $ km/s). Their column densities span the range 5$\times10^{18} {\rm cm}^{-2} - 5.7\times 10^{20}$ cm$^{-2}$. Some clouds exceed the threshold $N_{\ion{H}{I}} < 4 \times 10^{20} {\rm cm}^{-2}$ placed when defining the mask for the high latitude regions. This happens because some superpixels that lie at the edge of the mask contain high-resolution pixels with $N_{\ion{H}{I}}$ exceeding the applied threshold. There are 1348 superpixels which contain clouds with column densities higher than the column density threshold of the sky mask, which make up only 0.2\% of pixels in the studied sky area.

The highest column density clouds are found in the range of $\rm v_{LSR}$ that traditionally corresponds to LVCs \citep[$\pm 35$ km/s, chapter by Albert \& Danly in][]{HVCs2004}. Clouds within this velocity range exhibit similar ranges of $N_{\ion{H}{I}}$ in the North and South regions. At more negative velocity, cloud properties in the North and South regions differ: there are more IVCs at negative velocity with $N_{\ion{H}{I}} > 0.5\times 10^{20}$ cm$^{-2}$ in the North than in the South. The excess of IVC emission at negative velocities in the North is well-documented \citep[e.g.][]{HVCs2004} and has been known since early \ion{H}{I} surveys \citep[e.g.][]{Blaauw1966,wesselius1973}.

In the literature, the velocity that marks the boundary between LVCs and IVCs varies by tens of km/s \citep[see e.g.][where the cut is at $\pm 20$ km/s, $\pm 30$ km/s and $\pm 40$ km/s, repsectively]{magnani2010,wakker1991,wakker2001}. Here, we can use the second dimension of cloud $N_{\ion{H}{I}}$ to select a more suitable velocity cut for the selected sky regions. We will exploit the apparent similarity of LVC properties in the North and South.

For our selected regions, we expect LVCs in the North and South to have statistically similar physical properties and to reside within a few hundred parsecs of the Sun, based on several lines of evidence. First, the fact that all high-column density clouds are LVCs is consistent with the scale height of the \ion{H}{i} gas \citep{kalberla2007}. Second, linear features in \ion{H}{i} at low velocity are correlated with interstellar polarization of stars within a few hundred parsecs of the Sun \citep{clark2014}. Third, the majority of interstellar polarization arises just outside the Local Bubble at high latitude \citep{santos2011,berdyugin2014}, as does the majority of dust extinction detected in 3D dust maps \citep{lallement2019L}. In contrast, most negative velocity IVCs are more distant objects, associated with the Intermediate Velocity (IV) Arch, IV Spur, and other well-known complexes \citep{kunzdanly}. Constraints on the value of the distance of Northern IVCs from the Galactic midplane range widely \citep[e.g.][]{wakker2001,welsh2004,puspitarini2012}. Most sightlines have height brackets of 0.5-2 kpc \citep[]{wakker2001}. Exceptions exist for both cloud categories: extragalactic gas from parts of the Magellanic Stream is known to have low $\rm v_{LSR}$ \citep[e.g.][]{donghia2016}, while the molecular cloud IVC 135+54 \citep[IV21][]{kunzdanly} lies at a distance of $\sim$ 300 pc \citep{IVC135distance}, as do some intermediate-latitude parts of the IV Arch \citep{welsh2004}. Both these exceptions, however, only affect a small percentage of sightlines of the high latitude sky. Consequently, we expect that the physical properties of LVCs studied here differ statistically from those of IVCs.


The higher column density of LVCs and their expected similarity between hemispheres allows us to define a simple criterion for selecting the LVC velocity range. Figure \ref{fig:nhvel} (right), shows the 1- and 99-percentiles of the distribution of cloud velocities after selecting all clouds above a threshold in $N_{\ion{H}{i}}$. As the threshold is increased, we find improving agreement between the percentiles of the Northern and Southern $\rm v_{LSR}$ distributions. For $N_{\ion{H}{i}} > 2.5 \times 10^{20}$ cm$^{-2}$ the North and South velocity ranges are the same to within the first decimal. We thus adopt a velocity range for LVCs of $\rm -12 \, km/s < v_{LVC} < 10 \, km/s$. Clouds with centroid velocities outside this range fall in the IVC category.

\begin{figure*}
\includegraphics[scale=1]{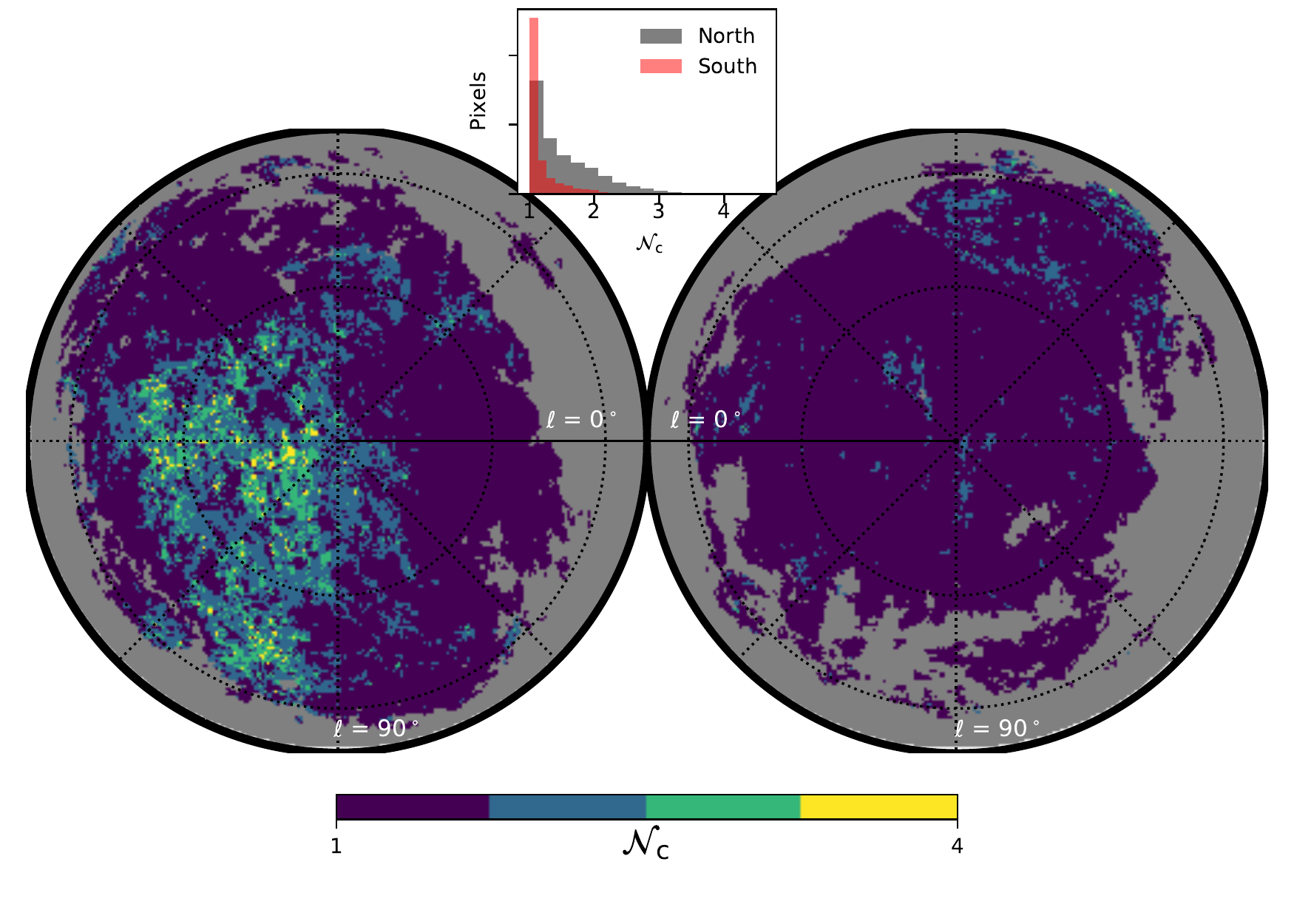}
\caption{Column-density-weighted number of clouds ($\rm \mathcal{N}_{c}$) per $N_{\rm{side}}$ 128 pixel in the North (left) and South (right) regions in the velocity range $|\rm v_{LSR}| < 70\, km/s$. The inset shows the 1D distribution of $\rm  \mathcal{N}_{c}$ per pixel for the North (gray) and South (red) regions.}
\label{fig:neff}
\end{figure*}

Note that placing a cut on the centroid velocity of clouds results in a more natural separation between LVCs and IVCs than applying a threshold in the \ion{H}{i} spectra, as is commonly done. This has the advantage of allowing for a gradual fall of the cloud signal, which can extend past the border in velocity. The emission of a cloud is thus not truncated arbitrarily.

Having defined a velocity cut to separate LVCs from IVCs, we examine how these classes are distributed on the sky. Figure \ref{fig:NcloudsLVCIVC} shows maps of $\rm N_{clouds}$ per pixel for the LVC and IVC range separately. The number of distinct kinematic components in the LVC range is significanlty smaller than that in the IVC range: 95\% of pixels have $\rm N_{clouds} \leq 1$ in the LVC range, a value that only 36\% of pixels in the IVC range have. LVCs cover almost the entire selected area, while IVCs are more clustered, and occupy a smaller sky fraction in the South than in the North. The patterns seen in Figure \ref{fig:nclouds_map} can be attributed primarily to IVCs, as can be seen by comparing with the lower left panel of Figure \ref{fig:NcloudsLVCIVC}. 

In the North, at longitudes $l$ = 90$^\circ$ $-$ 270$^\circ$, there are more clouds per pixel than in the rest of the area. This area is primarily occupied by negative velocity IVCs, that are associated with extraplanar gas structures, such as the IV Arch \citep{kunzdanly}. However, not all IVCs are extraplanar. Positive velocity IVCs at longitudes $l = 300^\circ - 350^\circ$ are likely associated with planar gas that does not lie in our immediate vicinity. In the South, there are pixels that contain emission from the Magellanic Stream, which coincides with low $\rm v_{LSR}$ and can easily be confused with Galactic emission. A more detailed analysis of the kinematics of the Galactic gas and the Stream is necessary in order to separate the two contributions \citep[e.g.][]{nidever}.

The fact that the IVC velocity range has a higher number of clouds per pixel than that of LVC gas can be due to the large distance to IVCs \citep[e.g.][]{wakker,wakker2008}. A pixel of 0.45$^\circ$ covers 8 pc in angular size at a distance of 1 kpc, and only 0.8 pc at 100 pc (nearest distance to Local Bubble wall). Moreover, it is likely that the majority of LVCs reside at the boundary of the Local Bubble, a structure with simple kinematics. At the same time, structures like the IV Arch may originate from a Galactic fountain process \citep{kunzdanly}, thus inheriting complex kinematics. However, another important factor is that the velocity range occupied by LVCs is only 20 km/s wide, much narrower than that of IVCs. At the selected KDE bandwidth of 5 km/s, our method cannot distinguish between velocity components that differ by less than $\sim$12 km/s. For the goal of separating IVCs from LVCs, the selected bandwidth is sufficient and necessary to avoid the detection of spurious peaks in the PDF of GC velocities (see Appendix \ref{sec:validation}). 

In our discussion so far, we have not introduced any weighting in counting the number of components along the line of sight. However, the column density of clouds varies both along the line of sight, as well as on the plane of the sky (Figure \ref{fig:NcloudsLVCIVC}, right panels). We define a measure of the complexity of the gas distribution along the line of sight that takes into account the cloud column density:
\begin{equation}
  \mathcal{N}_{c} = \sum^{\mathrm{N_{clouds}}}_{i=1} \frac{N^i_{\ion{H}{i}}}{N^{\rm max}_{\ion{H}{i}}},
  \label{eqn:Nc}
\end{equation}
where $N^i_{\ion{H}{i}}$ is the column density of the $i$-th cloud along the sightline (within the superpixel) and $N^{\rm max}_{\ion{H}{i}}$ is the column density of the cloud with highest $N_{\ion{H}{i}}$ in the superpixel. If a sightline has two clouds of equal column density, $\rm \mathcal{N}_{c}$ will be equal to 2. If one of two clouds has half the column density of the other, then $\rm \mathcal{N}_{c} = 1.5$, and so forth.

Figure \ref{fig:neff} shows maps of $\rm \mathcal{N}_{c}$ for the North and South regions. We again find an asymmetry between the two polar areas, with the North showing a higher fraction of pixels with $\rm \mathcal{N}_{c} > 1$. The 1D distributions of $\rm  \mathcal{N}_{c}$ differ significantly from those of $\rm N_{clouds}$ (Figure \ref{fig:nclouds_map}). In the North, the percentage of pixels with $\rm  \mathcal{N}_{c} < 1.5$, is 60\%, in the South it is 90\%. Thus, most sightlines are dominated by the column density of one cloud, with notable exceptions existing due to the presence of Northern IVCs. In the North, 17\% of pixels have $\mathcal{N}_c \geq 2$.


The distributions of $\mathcal{N}_c$ do not show tails extending to large values, unlike the distributions of $\rm N_{clouds}$. This stems from the fact that a large population of low column density clouds exist in our sample (see Figure \ref{fig:nhvel}).
Clouds with $N_{\ion{H}{i}} \lesssim 2\times 10^{19} \rm cm^{-2}$ may be affected by certain systematics, as discussed in Section \ref{ssec:limitations}. This population is most sensitive to the choice of superpixel size in the cloud identification step (Appendix \ref{sec:validation}). The column-density weighted number of clouds, $\mathcal{N}_c$, is insensitive to these problems. We show that the distribution of $\mathcal{N}_c$ remains stable for different choices of the superpixel size in Appendix \ref{sec:validation}. Compared to $\rm N_{clouds}$, $\mathcal{N}_{c}$ is a more robust measure of ISM complexity along the line of sight, in the context of CMB foreground subtraction.

\subsubsection{Signatures of IVCs in \textit{Planck} dust emission}
\label{ssec:ivc-correlations}

The previous Section shows that higher values of $\rm N_{clouds}$ and $\mathcal{N}_{c}$ are associated with the presence of IVCs. Dust properties have been found to differ between IVCs and LVCs in the 14 fields analysed by \citep{planck2011_xxiv}. We wish to investigate whether there are signs of varying dust properties associated with IVCs throughout the entire Northern high Galactic latitude sky. For this we do not attempt to repeat the joint analysis of \ion{H}{i} and \textit{Planck} dust emission presented in \citep{planck2011_xxiv}, as it is beyond the scope of the present work. Instead, we search for signatures of varying dust properties in the maps of dust model parameters produced by \cite{PlanckGNILC2016}. 

We use the map of dust temperature, $T_d$, created by fitting a modified black body to multi-frequency dust emission maps\footnote{From the Planck Legacy Archive: \texttt{COM\_CompMap\_Dust-GNILC-Model-Temperature\_2048\_R2.01.fits}}. The \textit{Planck} and IRIS data were processed with the Generalized Needlet Internal Linear Combination (GNILC) algorithm \citep{remazeilles2011} to separate the contribution of the cosmic infrared background to the observed emission. We also make use of the $\chi^2$ map that resulted from the model fit to GNILC-processed dust emission maps \citep{PlanckGNILC2016}\footnote{M. Remazeilles, private communication.}. We downgrade both maps from their native resolution of $N_{\rm{side}} = 2048$ to $N_{\rm{side}}$=1024, to match the resolution of the HI4PI data. The morphological patterns that appear on the maps under investigation typically cover sky areas much larger than the pixel size at any of these resolutions. Averaging model parameters in this way is only used here for investigating such large-scale spatial correlations \citep[see also][]{hensley2019}.

We create maps of \ion{H}{i} column density of LVCs and IVCs separately at a resolution of $N_{\rm{side}}$=1024 (as mentioned in Section \ref{sec:method}). They are used to make a map of the ratio, $\rho$, of IVC to LVC column density:
\begin{equation}
\rho = \frac{N^{\rm IVC}_{\ion{H}{i}}}{N^{\rm LVC}_{\ion{H}{i}}}.
\end{equation}
In pixels containing an LVC but no IVC, we set $\rho = 10^{-3}$. In pixels where only IVCs were detected, we set $\rho = 10^3$. Masking out these pixels instead has no effect on the result. Pixels where neither IVCs nor LVCs are found are masked. We apply this mask to the $T_d$ map as well.

\begin{figure*}
\includegraphics[scale=1]{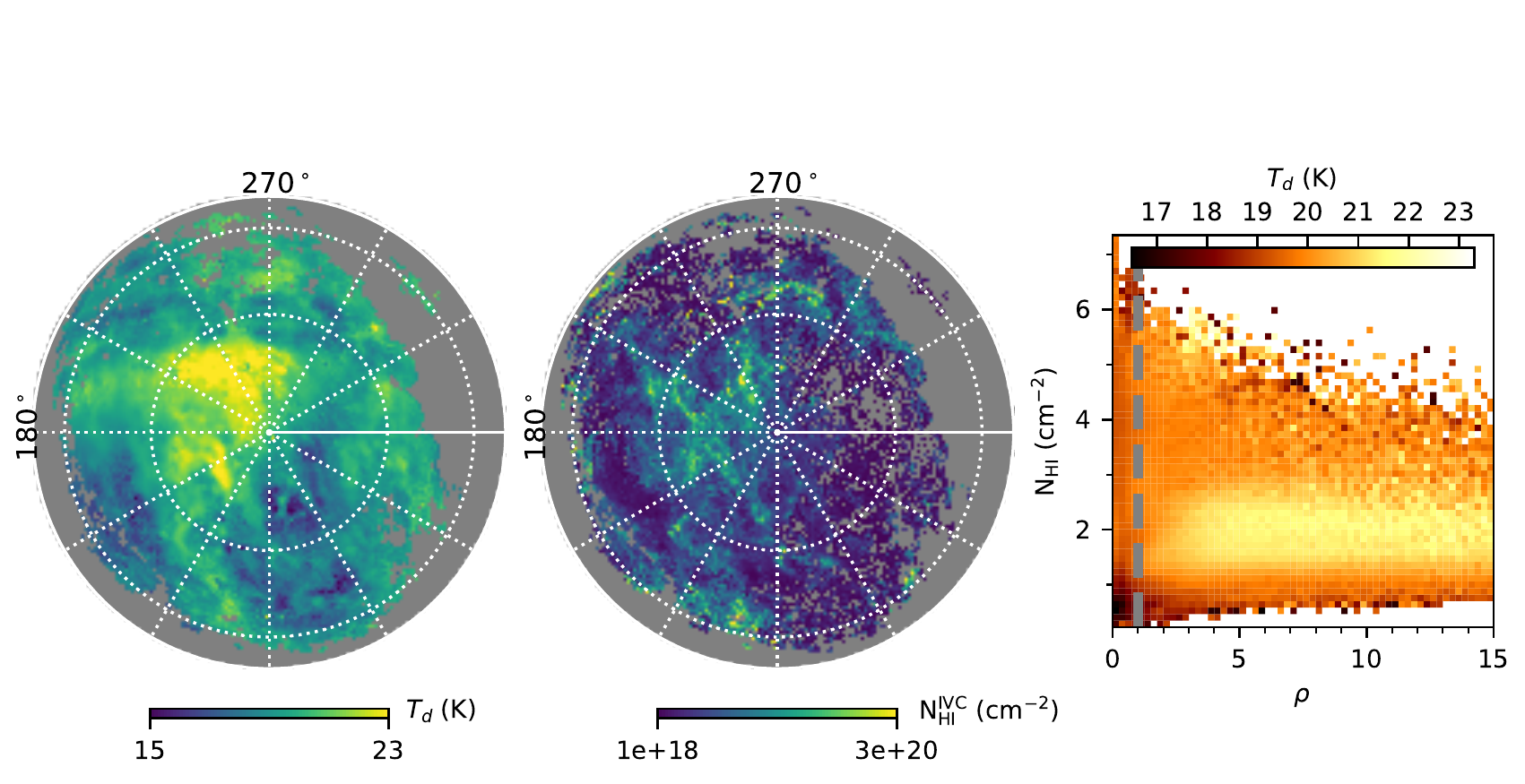}
\caption{\textit{Effect of IVCs on derived dust temperature:} Morphological connection between maps of dust temperature, $T_d$ (left), and IVC column density $N^{\rm IVC}_{\ion{H}{i}}$ (middle). Right: 2d distribution of total $\rm N_{\ion{H}{i}}$ versus the ratio $\rho$ of IVC to LVC column density. The color shows the mean $T_d$ in each bin. The vertical dashed line marks the value $\rho=1$, on either side of which we find significantly different $T_d$, for given $\rm N_{\ion{H}{i}}$. The range of $\rho$ shown is truncated to [0,15] for better visualization.}
\label{fig:Temp_NH_r}
\end{figure*}

Figure \ref{fig:Temp_NH_r} shows the maps of $T_d$ and IVC column density $N^{\rm IVC}_{\ion{H}{i}}$. The $T_d$ map shows large-scale, coherent variations with a significant enhancement of $T_d$ near the Pole \citep[as noted in][]{planckXI2014}. There is an apparent tendency for pixels with higher $N^{\rm IVC}_{\ion{H}{i}}$ to show higher $T_d$, though the correspondence is not one-to-one. This morphological similarity was found in \cite{planckXI2014}, and remains in our maps despite the different methods of creating the $N^{\rm IVC}_{\ion{H}{i}}$ map and the different $\rm v_{LSR}$ cut used to separate LVCs and IVCs ($\pm 35$ km/s was used in their work).

A simple correlation of $T_d$ and $N^{\rm IVC}_{\ion{H}{i}}$ yields the very small Pearson correlation coefficient of 0.2. However, we need to take into account that $T_d$ is also found to correlate with the total $\rm N_{\ion{H}{i}}$, as discussed by \cite{hensley2019}. If IVC intrinsic properties are responsible for the change in $T_d$, then we should expect that sightlines with stronger IVC emission show a higher $T_d$ when compared to LVC-dominated regions \textit{with the same total} $\rm N_{\ion{H}{i}}$. This is indeed what we find, as demonstrated in Figure \ref{fig:Temp_NH_r} (right panel). By binning pixels in the $\rm N_{\ion{H}{i}}$ $-$ $\rho$ plane, we find that the average $T_d$ for pixels with $\rho>1$ is higher than that for $\rho<1$. For the same range of $\rm N_{\ion{H}{i}}$, namely [1.0, 2.8]$\times 10^{20}$ cm$^{-2}$, the mean $T_d$ changes from 19.4 K to 20.4 K as we transition from sightlines dominated by LVCs ($\rho<1$) to those dominated by IVCs ($\rho>1$). Beyond $\rho=1$ there is very little variation in $T_d$. The mean $T_d$ remains higher for  $\rho>1$ compared to $\rho<1$ even if we do not restrict $\rm N_{\ion{H}{i}}$ to the aforementioned range (in which case we find a mean $T_d$ of 20.1 and 19.3, respectively).

We hypothesize that the change in $T_d$ is most prominent for a restricted range of $N_{\ion{H}{i}}$ because that is the range where negative velocity IVCs dominate. At lower latitude and higher column, positive velocity IVCs dominate, and these are most likely not extraplanar gas. To test this, we select pixels where the $ N_{\ion{H}{i}}$ from negative velocity, $N^{\rm IVCneg}_{\ion{H}{i}}$, is at least half of the total IVC $N_{\ion{H}{i}}$. We find that 54\% of pixels that satisfy this condition have IVC $N_{\ion{H}{i}}$ in the range [1.0, 2.8]$\times 10^{20}$ cm$^{-2}$. If we look at pixels that also have $\rho>1$, we find that a significantly larger percentage of those (74\%) lies in the aforementioned column density range. Therefore, it is likely that the negative velocity IVCs are those responsible for the increase in $T_d$ within the column density range [1.0, 2.8]$\times 10^{20}$ cm$^{-2}$.

\begin{figure*}
\includegraphics[scale=1]{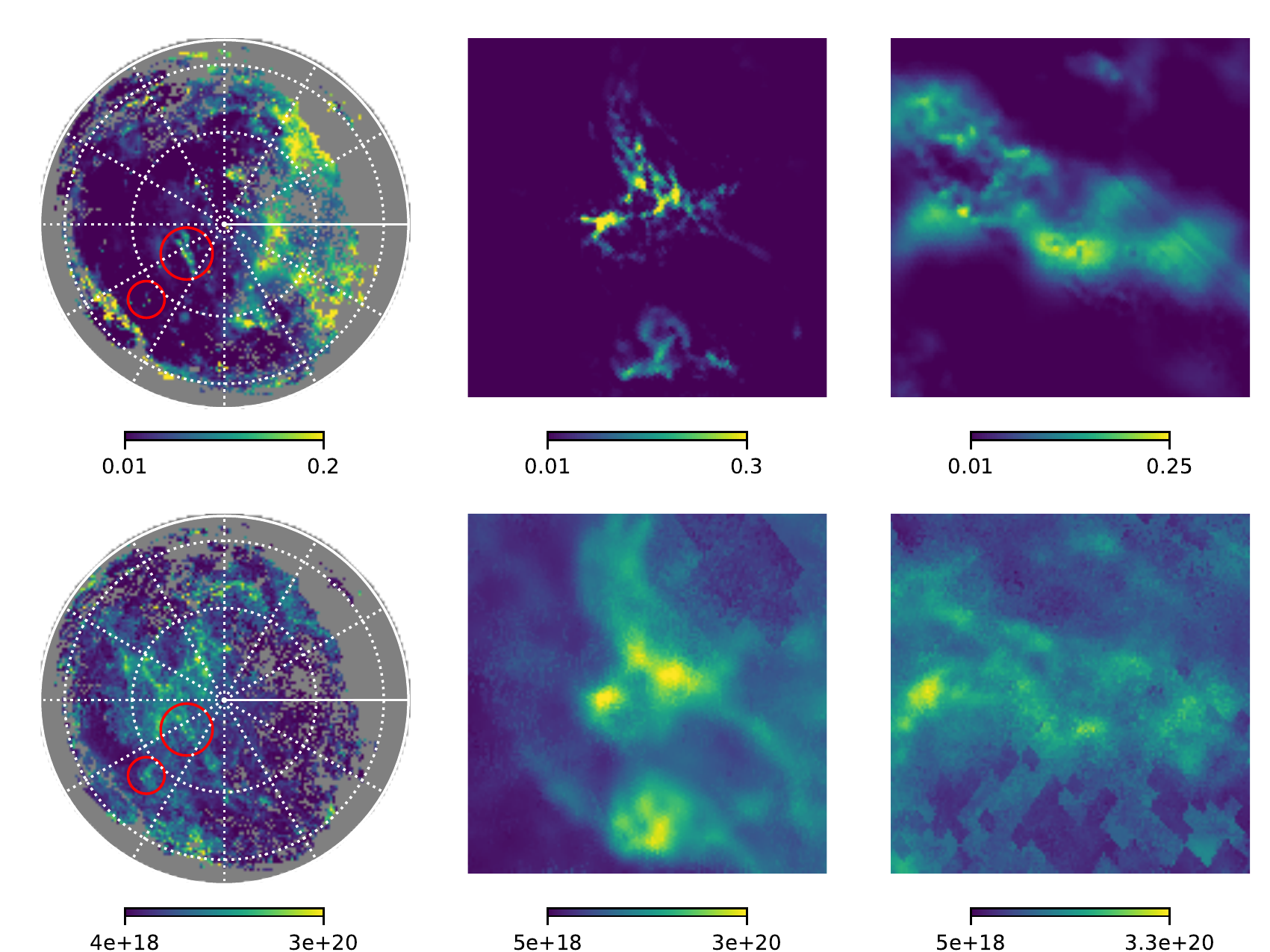}
\caption{\textit{Morphological correlation between $\chi^2$ (top row) and $N^{\rm IVC}_{\ion{H}{i}}$ (bottom row).} From left to right: maps showing the Northern Galactic Polar region, zoom-ins centered on $l, \,b$ = (136$^\circ$, 54$^\circ$) and (142$^\circ$, 75$^\circ$) (image sizes are 7$^\circ$ and 10$^\circ$, respectively).}
\label{fig:chi2}
\end{figure*}

The evidence that dust in IVCs is different than that in LVCs implies that the dust Spectral Energy Distribution (SED) might be more complex than a modified black body along sightlines where both classes of cloud exist. If deviations from a modified black body SED exist in such sightlines, then the model used in \cite{planckXI2014} might yield slightly poorer fit than the rest of the high-latitude sky. We investigate if this is the case by using the $\chi^2$ statistic (Figure \ref{fig:chi2}). 

A correlation with $N^{\rm IVC}_{\ion{H}{i}}$ is not as obvious in the $\chi^2$ map as it was in the $T_d$ map. However, we do find certain regions where the $\chi^2$ is consistently elevated with patterns that morphologicaly match certain IVCs. Figure \ref{fig:chi2} shows zoomed-in portions of the $\chi^2$ and $N^{\rm IVC}_{\ion{H}{i}}$ maps centered on prominent features in the former. The cloud shown in the middle panels is the well-studied diffuse molecular cloud IVC135+54-45 (IV21) \citep{kunzdanly,IVC135distance,LenzIVC2015}, known to have lower metallicity than LVC gas \citep{IVC135_FIR}. The feature on the right panels extends over $\sim 10^\circ$, is part of the IV Arch and overlaps with clouds IV2,4,11,17 in the catalog by \cite{kunzdanly}. We also find close morphological matches between $\chi^2$ and $N^{\rm IVC}_{\ion{H}{i}}$ for other known molecular IVCs that are less extended on the sky. We visually inspected the maps of $\chi^2$ and $N^{\rm IVC}_{\ion{H}{i}}$ at the locations of the MIVCs in the catalog of \cite{mivcs2016} (within an area of 4$^\circ \times$4$^\circ$). We found 36 locations with clear similarities, which amounts to 40\% of MIVCs within the footprint of our $ N^{\rm IVC}_{\ion{H}{i}}$ map.

We note that our high-resolution $N^{\rm IVC}_{\ion{H}{i}}$ maps contain evident artifacts of the pixelization used in the cloud identification method. These arise from the IVC/LVC selection and are largely absent in the total $\rm N_{\ion{H}{I}}$ map. We discuss this artifact in Section \ref{sec:discussion}. For the purpose of quantifying the number of clouds along the line of sight (relevant to CMB foreground modeling), the statistics of individual sightlines are the primary target. While a more advanced method that includes spatial correlations on scales larger than the superpixel size should be pursued in order to resolve such artifacts, we do not expect any significant effect on the statistics of $\rm N_{clouds}$ or $\mathcal{N}_{c}$.

So far, we have investigated the effect of IVCs on the total intensity of dust emission. We briefly examine whether IVCs leave a traceable imprint on the polarized dust emission measured by \textit{Planck} at 353 GHz. We follow \cite{planckstars} to construct smooth maps of Stokes I, Q and U at a resolution of 80\arcmin. We subtract the zero-level offset of 389 $\mu K_{CMB}$ from the I map. After smoothing, the map resolution is downgraded to $N_{\rm side} = 128$. We construct a map of fractional linear polarization, $p_{353}$, by using the equation $p_{353} = \sqrt{Q^2+U^2}/I$. 

\begin{figure}
\includegraphics[scale=1]{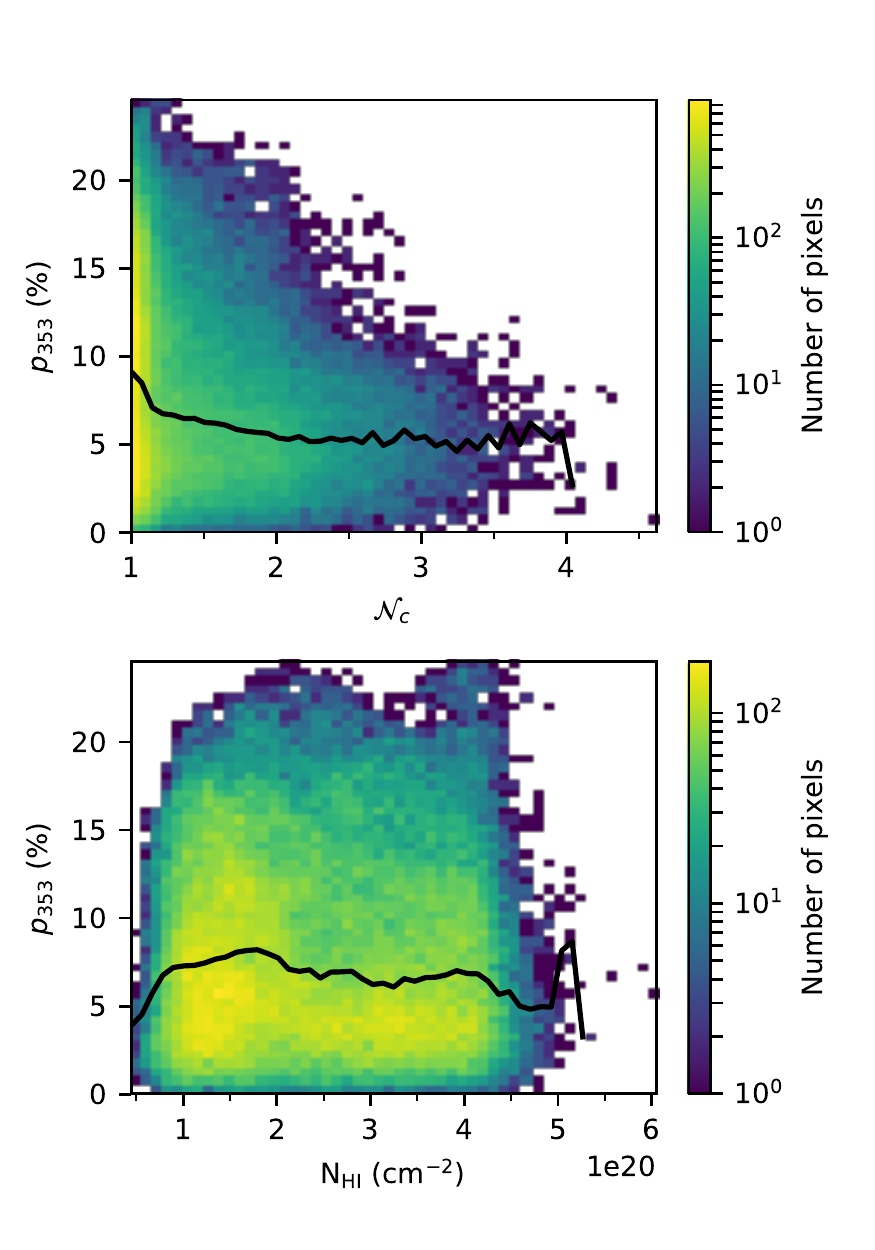}
\caption{\textit{Effect of IVCs on polarized dust emission.} Top: The joint distribution of fractional linear polarization at 353 GHz, $p_{353}$, and $\mathcal{N}_c$. Bottom: The joint distribution of $p_{353}$ and $\rm N_{\ion{H}{i}}$. A black line marks the running median of each distribution.   }
\label{fig:p-ivc}
\end{figure}

Figure \ref{fig:p-ivc} (top) shows the joint distribution of $p_{353}$ and the column-density-weighted number of clouds per pixel ($\mathcal{N}_c$) from Section \ref{ssec:Nclouds}. Intriguingly, there is a marked decrease of the median and maximum $p_{353}$ at high $\mathcal{N}_c$. As discussed in Section \ref{ssec:Nclouds}, larger values of $\mathcal{N}_c$ are found primarily in regions where IVCs are prominent. The decrease of $p_{353}$ with $\mathcal{N}_c$ may indicate some level of line-of-sight depolarization caused by IVCs. Note that we have not taken into account the uncertainties in $p_{353}$, which may cause positive bias for pixels with low signal-to-noise ratio. The trend shown in Figure \ref{fig:p-ivc} (top) is distinct from the decrease in $p_{353}$ with total $\rm N_{\ion{H}{i}}$ found at higher column density by \cite{planckstars}. In the high-galactic latitude regions studied here, $p_{353}$ is primarily flat with $\rm N_{\ion{H}{i}}$ (Fig. \ref{fig:p-ivc}, bottom). We find that $p_{353}$ is related to the IVC column density, rather than the total column density: a trend similar to that in Fig. \ref{fig:p-ivc} (top) is seen also when comparing with IVC $\rm N_{\ion{H}{i}}$ instead of $\mathcal{N}_c$. 

Together, the results of this Section indicate that IVCs are likely an important contributor to the \textit{Planck} dust emission, both in total and in polarized intensity.

\subsection{The BICEP/Keck field}
\label{sec:bicep}

A primary motivation for this work is to inform CMB foreground modeling efforts. To this end, we focus on the region targeted by the BICEP/Keck experiment \citep{BICEP1}. We examine the region as defined by the mask provided on the experiment's website \footnote{http://bicepkeck.org}, after downgrading from $N_{\rm{side}}$ = 512 to 128. We use the method outlined in Section \ref{sec:method} to quantify the complexity of $\rm \ion{H}{I}$ spectra. As in Section \ref{ssec:Nclouds}, we analyse the velocity range $\rm |v_{LSR}| \leq 70$ km/s to avoid most HVC emission. We note that parts of the Magellanic Stream in this area have $\rm |v_{LSR}|$ $\sim$ 70 km/s \citep{westmeier}, and are therefore not removed by this cut.

\begin{figure*}
\includegraphics[scale=1]{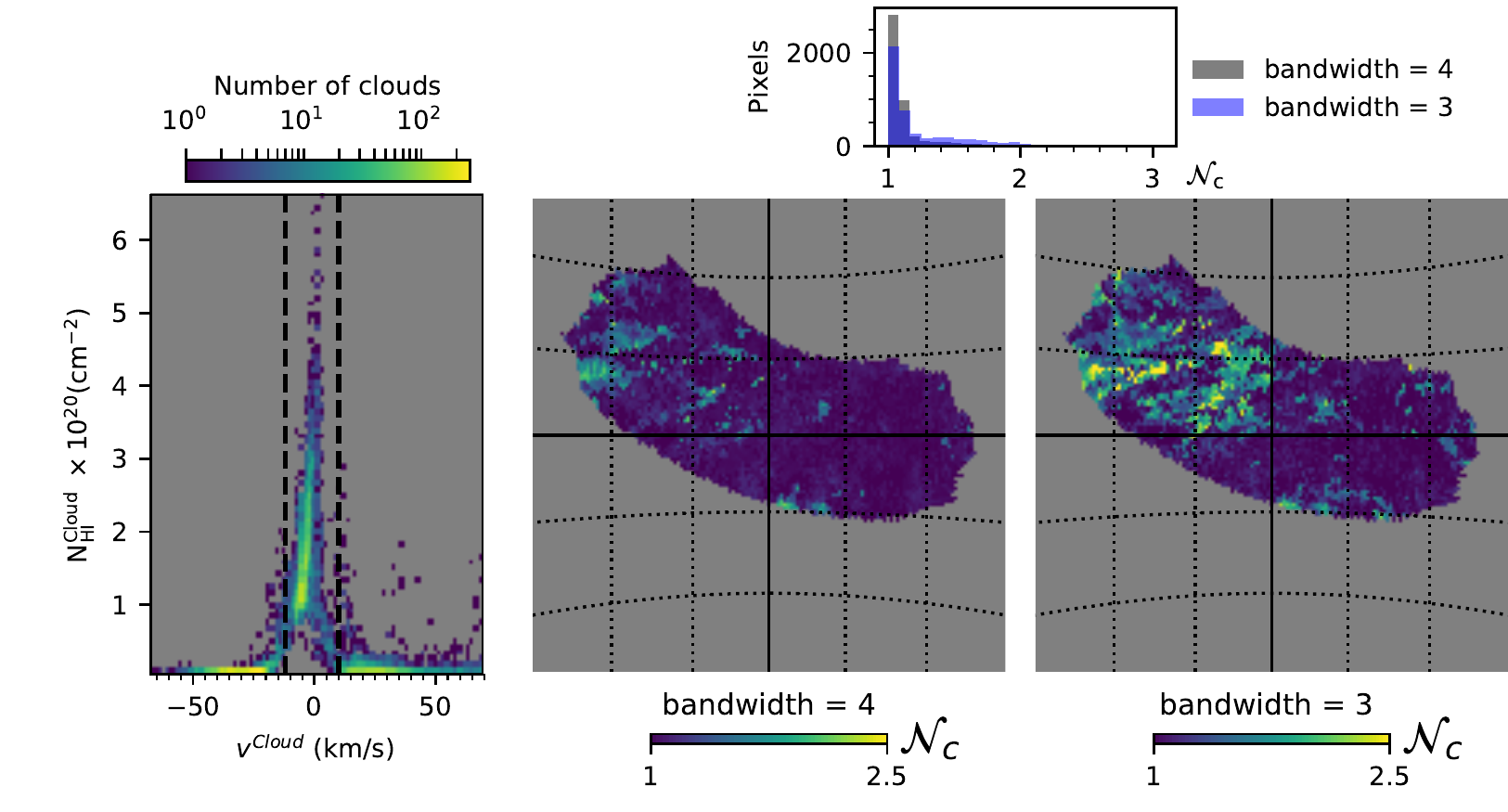}
\caption{\textit{The number of clouds per pixel in the BICEP/Keck region.} Left: Two dimensional distribution of cloud $\rm N_{\ion{H}{i}}$ and cloud centroid velocity (from equation \ref{eqn:vcentroid}). Middle: Map of the column-density-weighted number of clouds, $\mathcal{N}_c$, for a KDE bandwidth of 4 channels. Right: $\mathcal{N}_c$, for a KDE bandwidth of 3 channels. Top inset shows the distribution of $\mathcal{N}_c$ for both choices of bandwidth. Grid lines are spaced by 10 degrees.}
\label{fig:bicep_Nc}
\end{figure*}

The dominant cloud in each pixel lies within the LVC range, as can be seen by examining the joint distribution of cloud column density and cloud centroid velocity (Fig. \ref{fig:bicep_Nc}, left). The majority of high-$\rm N_{\ion{H}{i}}$ clouds are found at low velocity, in agreement with the results presented previously for larger sky regions. Clouds with very low $\rm N_{\ion{H}{i}}$ ($N_{\ion{H}{i}} < 0.5\times10^{20} {\rm cm}^{-2}$) are found at intermediate velocities. The tail of low-$\rm N_{\ion{H}{i}}$ clouds that extends to high (positive) velocity is associated with the Magellanic Stream, and thus will likely not contain traceable amounts of dust. 

Figure \ref{fig:bicep_Nc} (middle) shows the column-density-weighted number of clouds per pixel, $\mathcal{N}_c$. $\mathcal{N}_c$ ranges from 1 to 2.3 with 75\% of pixels having $\mathcal{N}_c < 1.1$ (for a 2-cloud sightline this would imply that one cloud has a tenth of the column density of the other). Thus, one cloud dominates the column density in the majority of pixels in the BICEP/Keck region. We calculate the relative contribution of the highest-$\rm N_{\ion{H}{i}}$ cloud per pixel compared to the total $\rm N_{\ion{H}{i}}$ of the sightline. We find that for 85\% of pixels the highest-$\rm N_{\ion{H}{i}}$ cloud contributes more than 80\% of the column.

We note that our definition of a cloud depends on the choice of KDE bandwidth. The method is not able to distinguish between velocity components that differ by less than $\sim$2.5 times the chosen bandwidth. Therefore, it is possible that within the LVC range there may exist kinematicaly distinct features that are identified as a single `cloud'. For a bandwidth of 4 channels (5 km/s) we find that there is unresolved substructure in the identified clouds. We quantify the amount of substructure by identifying maxima in the mean spectrum of each cloud. Two or more maxima are found for 40\% of LVCs in the region. Thus, while IVCs are likely not a concern for foreground modeling in this region, the LVC substructure  may be cause to consider models with multiple dust components. The very nature of the multi-phase neutral medium may cause such velocity substructure, as well as variations in dust properties \citep[e.g.][]{clark2019,murray2020}, motivating the incorporation of phase structure into foreground modeling \citep{ghosh2017,adak2019}.

We further investigate the velocity substructure by varying the choice of bandwidth. We repeat the analysis by choosing a bandwidth of 3 channels (3.8 km/s) and show the resulting $\mathcal{N}_c$ in the right panel of Figure \ref{fig:bicep_Nc}. Much of the substructure in the cloud spectra is now resolved into separate clouds. $\mathcal{N}_c$ has a maximum value of 3.0 with 54\% of pixels having $\mathcal{N}_c < 1.1$. Pixels with $\mathcal{N}_c > 1.5$ take up 20\% of the area, while those with $\mathcal{N}_c > 2$ amount to only 4\%. The fraction of pixels where the highest-$\rm N_{\ion{H}{i}}$ cloud contributes more than 80\% of the column is now 61\%. The conclusion remains that $\rm N_{\ion{H}{i}}$ is dominated by one cloud per superpixel for most of the area (within $\rm |v_{LSR}| \leq 70$ km/s).

\begin{figure}
\centering
\includegraphics[scale=1]{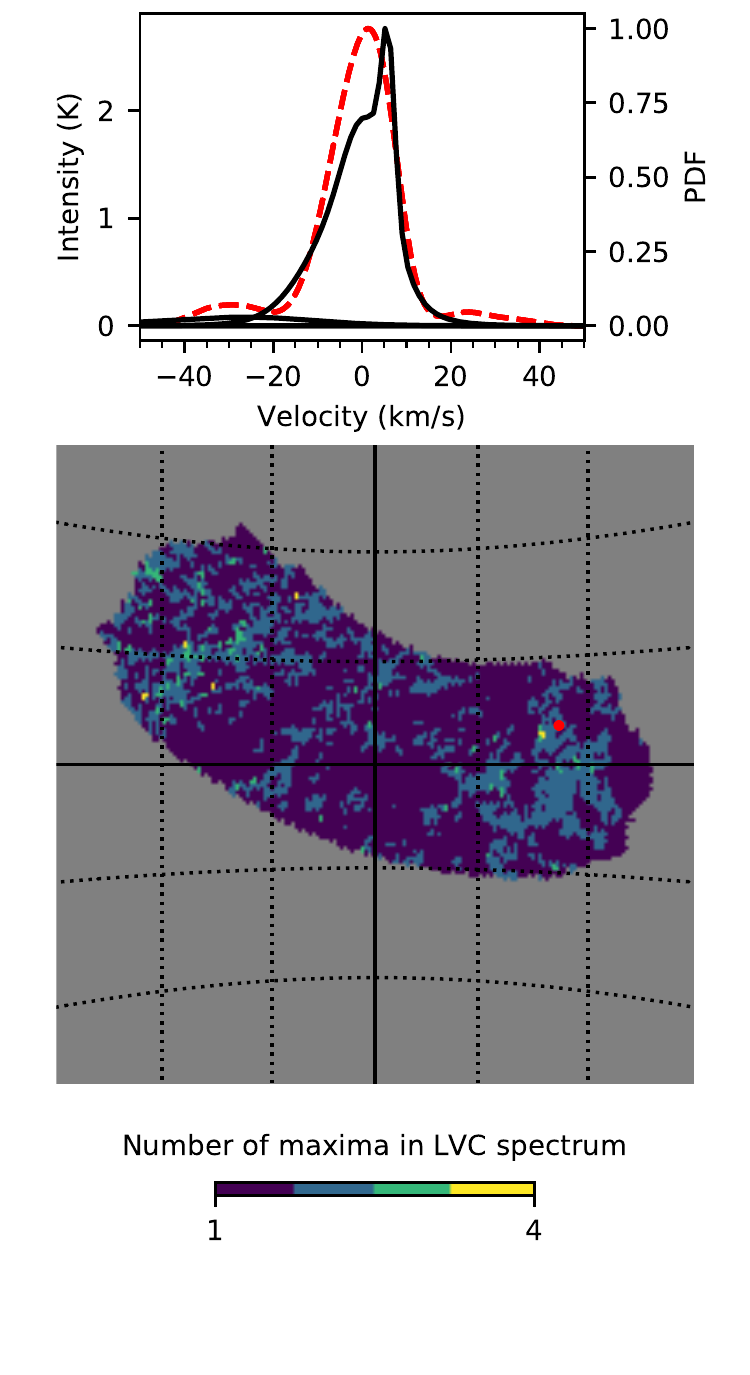}
\caption{\textit{Substructure in the mean spectrum of LVCs}. Top: The red dashed line (right axis) shows the PDF of GC mean velocity of a superpixel in the region. The black lines (left axis) show the average spectrum of clouds in the superpixel (at each velocity we find the average intensity from the collection of all GCs of the cloud). The pixel shown is centered at (l,b) = (285.6$^\circ$, -52.4$^\circ$). The mean spectrum that peaks at $\sim 5$ km/s has two maxima, revealing structures unresolved by the cloud identification. Bottom: Map of the number of maxima in LVCs identified in the region. A red dot marks the pixel shown in the top panel. A KDE bandwidth of 3 channels was used.}
\label{fig:spectrabicep}
\end{figure}

Substructure in the mean spectrum of clouds exists even after using the higher velocity resolution (bandwidth = 3 channels). An example is shown in Fig. \ref{fig:spectrabicep} (top), where two maxima are seen in the mean spectrum of an LVC. To quantify this effect, we construct a map of the number of maxima per cloud. In each pixel we identify the highest column density cloud in the LVC range and measure the number of maxima in its mean spectrum. The resulting map is shown in Figure \ref{fig:spectrabicep} (bottom). We find spatially coherent regions with $\sim$ 2 maxima, primarily in the Western and Eastern areas of the map. Two or more maxima are found in 30\% of pixels.

\begin{figure*}
\centering
\includegraphics[scale=1]{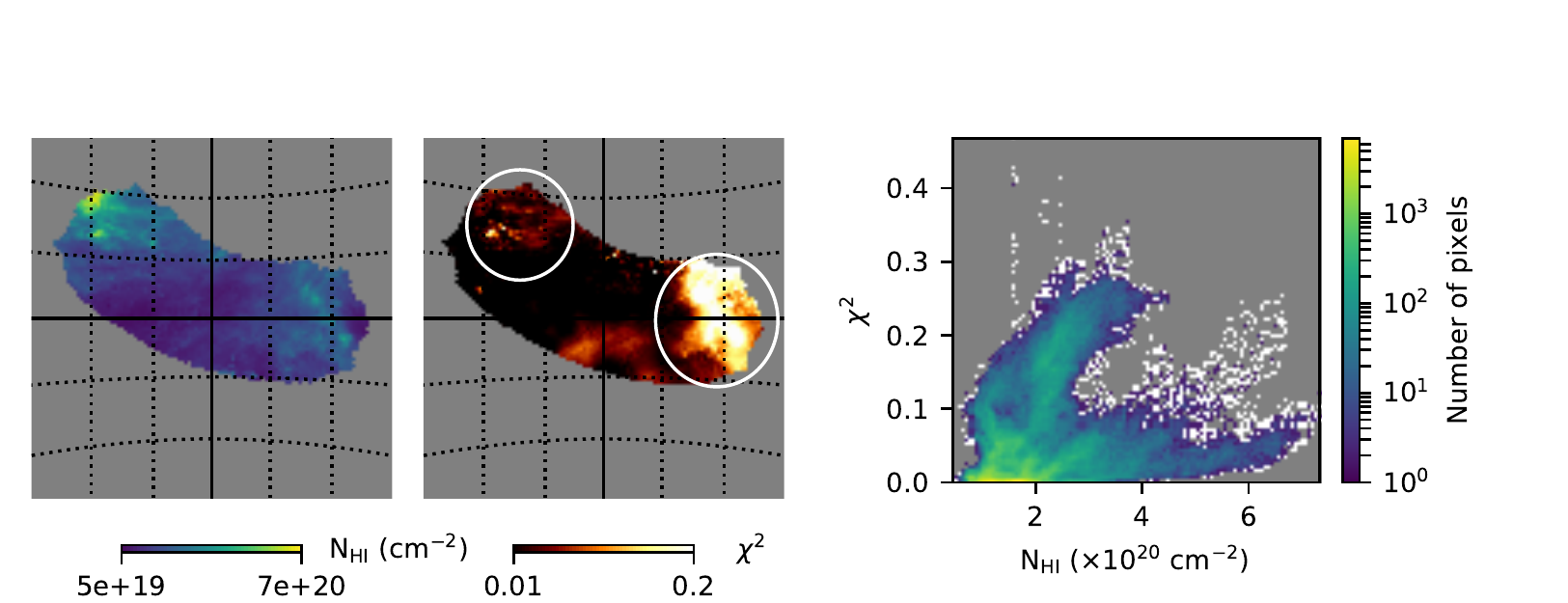}
\caption{Maps of $N_{\ion{H}{I}}$ (left) and dust-emission $\chi^2$ (middle) in the BICEP/Keck field and their two-dimensional distribution (right). In the middle panel, white lines encircle two regions that show spatial correlation between the two maps. The highly correlated population of pixels with $\chi^2 > 0.1$ at the low-$N_{\ion{H}{I}}$ range (right panel) corresponds to the Eastern circled region in the middle panel. The more modest correlation seen between $N_{\ion{H}{I}}$ of 4 and 7 $\rm \times 10^{20} cm^{-2}$ corresponds to the Western marked region in the middle panel.}
\label{fig:bicep_NH_chi}
\end{figure*}

In Section \ref{ssec:ivc-correlations} we found that regions containing multiple clouds (IVCs and LVCs) sometimes showed elevated values of $\chi^2$ of the dust SED model fit. With the absence of IVCs in the region, it is interesting to investigate if there are any morphological correlations between the $\chi^2$ map and the complexity of \ion{H}{i} spectra. The distribution of $\chi^2$ on the sky does show morphological similarities with the total column density map (middle and left panels of Fig. \ref{fig:bicep_NH_chi}, respectively). In particular, there are two regions of elevated $\chi^2$ that correlate with the \ion{H}{I} column. These are marked with white circles in the middle panel of Fig. \ref{fig:bicep_NH_chi}. The lower part of the map shows a ring feature that is not related to the \ion{H}{I} morphology. The right panel of Fig. \ref{fig:bicep_NH_chi} shows the 2D distribution of $\chi^2$ and $N_{\ion{H}{I}}$. The two marked regions correspond to two correlated regions in this plot. The Eastern region shows higher values of $\chi^2$ ($\chi^2 > 0.1$) but relatively modest $N_{\ion{H}{I}}$. The Western region contributes to the correlation seen at $ 4 \times 10^{20} {\rm cm}^{-2} < N_{\ion{H}{I}}<7 \times 10^{20} {\rm cm}^{-2}$. We separate the two regions with a simple cut on galactic longitude of $l = 315^\circ$. Within these sub-regions we find a Pearson correlation coefficient of 0.68 for the Eastern and 0.40 for the Western region. 

In Section \ref{ssec:ivc-correlations} we found that the existence of IVCs is sometimes correlated with elevated values of $\chi^2$. The origin of the spatially coherent high-$\chi^2$ values in the BICEP/Keck region, however, must be different, as there is no significant contribution of IVCs to the total \ion{H}{I} column density. Therefore, if the increased $\chi^2$ in this region is due to the presence of multiple dust components, then these components would correspond to clouds in the LVC range. If that is the case, we expect that LVCs in the regions of elevated $\chi^2$ will show more complex kinematics. The Western region does indeed show elevated values of $\mathcal{N}_c$ (Figure \ref{fig:bicep_Nc}) as well as LVC spectra with multiple components (Figure \ref{fig:spectrabicep}). The Eastern region, which has the highest $\chi^2$ values, does not appear to have elevated $\mathcal{N}_c$. However, this region shows a larger proportion of spectra with multiple (unresolved) components (Figure \ref{fig:spectrabicep}). We select pixels in this region by imposing $\chi^2 > 0.1$. We find that 38\% of these pixels have at least two maxima in the mean spectrum of clouds, compared to 26\% in the rest of the map. 

The regions that exhibit higher $\chi^2$ fall in the outskirts of the BICEP/Keck region and are assigned lower weights than the central pixels in the power spectrum analysis of \cite{BICEP2}. We isolate pixels where the mask values (weights) are higher than 0.9 and investigate the complexity of their associated \ion{H}{I} emission. We find that in 15\% of pixels the mean spectrum of the LVC component shows two maxima. These components may not exhibit as stark differences in dust properties as seen between LVCs and IVCs, as the $\chi^2$ is low throughout this central region of the BICEP/Keck field.

The required precision with which dust must be modeled as a CMB foreground depends on the target sensitivity of the experiments. Further work is needed to assess the effect of multiple clouds on the polarized dust SED in the region. The results presented in this Section can aid future investigations by providing estimates of the expected dust emission signal from each cloud along the line-of-sight.

\section{Discussion}
\label{sec:discussion}

By analysing the \ion{H}{I} emission at high Galactic latitude, we have produced a measure of the number of discrete components per line-of-sight that may contribute to the observed dust emission at microwave frequencies. Our primary motivation was to inform CMB-foreground modeling. We discuss our results in the context of the latest literature that aims to quantify line-of-sight effects in the ISM relative to CMB-foreground subtraction.

\subsection{Comparison of the number of clouds per sightline with previous works}

In recent years it has been realized that a major uncertainty in CMB foreground modeling comes from the unknown complexities of the dust distribution along the line-of-sight \citep[e.g.][]{TassisPavlidou2015}. Various estimates of the number of dust components along the line of sight have been used in order to understand the effect of such complexities on the recovery of cosmological parameters. The estimates differ drastically in the literature. In one of the first works assessing such effects, \cite{Poh2017} studied the region around the North Galactic Pole. They used two models: one with 9 clouds per kpc and another based on the 3D dust extinction map by \cite{green2015}. \cite{plancklayers} use a model of discrete layers of dust to describe the dust emission in the South Galactic Cap ($b < -60$).  They find that 4-9 layers can reproduce the 1-point statistics of the polarized dust emission from \textit{Planck}. In their approach, the number of layers was simply a free parameter of the model, unrelated to a specific physical quantity. In contrast, the model of \cite{ghosh2017} and \cite{adak2019} employs three discrete layers of dust, each associated with a different phase of the ISM. Their analysis is based on HI4PI data in the Galactic polar caps and also succeeds in reproducing observables in the polarized dust emission. The all-sky model of \cite{martinez-solaeche2018} is based on the 3D reddening map of \cite{green2015}. Dust emission at high Galactic latitudes in their map arises from $\sim$ 1-2 layers.

The variety of assumptions in such models was a prime motivator for the present work. Our measure of the number of \ion{H}{i} clouds along the line of sight shows that on average 2.5 (South) and 3 (North) kinematicaly distinct components are found at high Galactic latitude (Section \ref{ssec:Nclouds}). The relative contribution of these components along the sightline varies from pixel to pixel. We introduced a separate measure of the number of clouds that takes into account the relative column density of clouds. The Southern Galactic Polar region is dominated by the column density of one cloud per sightline. In contrast, the dominant cloud contributes less than 2/3 of the total column density for 40\% of Northern pixels (where $\mathcal{N}_c > 1.5$).

Our determination of the average number of clouds per pixel differs drastically from that assumed in some of the first works mentioned previously \citep{Poh2017,plancklayers}. These large numbers of 'layers' have been reduced in the physically-motivated models of \cite{ghosh2017}, \cite{adak2019} and \cite{martinez-solaeche2018}. 

When assuming three dust components, each associated with a discrete phase of the neutral ISM\footnote{(Cold Neutral Medium, CNM, Warm Neutral Medium, WNM, and Unstable Neutral Medium, UNM)}, \cite{adak2019} find that the Northern Galactic Polar region model lacks a necessary fourth component to account for all the observed dust. This fourth component corresponds to IVCs. In our determination of the number of clouds per pixel we do not distinguish between ISM phases, but find that there are on average 3 kinematicaly distinct components along the line of sight. Given the wide bandwidth chosen for the application of our method, it is natural that these components correspond to discrete multi-phase clouds. An improved description for dust modeling would combine the two approaches and take into account both the ISM phase information but also the discrete nature of LVCs and IVCs \citep[see also][]{murray2020}.

Most recently, a different class of models has been developed in which \ion{H}{I} velocity channel maps are treated as 'layers' of emission along the line of sight \citep{clarkhensley,hu2020,Lu2019}. These approaches utilize morphological information as well as the column density to successfully reconstruct properties of the polarized dust emission that are related to the 3D structure of the ISM magnetic field. To fully model the dust emission Stokes parameters, these approaches must make assumptions on the temperature and spectral index of dust within each 'layer'. These assumptions can be informed by the results presented in Section \ref{sec:results}, in particular regarding the constraints on differences between dust in IVCs and LVCs. 

\subsection{Are IVCs important for polarized CMB foreground subtraction?}

At high Galactic latitude the observed \ion{H}{I} column density that is correlated with reddening originates from LVC and IVC gas \citep[e.g.][]{planckXI2014,lenz2017}. These two classes of clouds may exhibit systematic differences in their dust and magnetic field properties, potentially causing variations of the polarization angle of dust emission with frequency \citep[as proposed by][]{TassisPavlidou2015}.

Evidence for differences in the dust properties of IVCs compared to LVCs was found when analyzing the total intensity of dust emission by \cite{planck2011_xxiv}. The increased dust temperatures and lower dust emission cross-sections in IVCs were attributed to smaller grain sizes in IVCs arising from dust shattering in the Galactic halo. This interpretation was also considered by \cite{planckXI2014} to explain the lower dust specific luminosity found in Galactic Polar regions with IVCs. In this work we noted two additional pieces of evidence that point to the different dust properties of IVCs compared to LVCs. One is a morphological correlation between the column density of some IVCs with the map of reduced $\chi^2$ from the modified black body fit to the dust emission SED performed by \cite{PlanckGNILC2016}. The elevated values of $\chi^2$ in regions with (primarily molecular) IVCs imply that the single-component modified black body yields a significantly poorer fit to the dust SED. The second is a systematic increase in the dust temperature fit parameter, $T_d$, for pixels where IVCs contribute more to the column density than LVCs. This correlation is true for \textit{a given column density}. The correlation is therefore not driven by the fitting degeneracy between $T_d$ and dust amplitude at low signal-to-noise ratio regions that was discussed in \cite{hensley2019}.

Some evidence for differences in magnetic field properties between IVCs and LVCs exists in the literature. To the best of our knowledge there are two indications of such variations based on analysis of starlight polarization. \cite{clark2014} note a possible loss of alignment between starlight polarization and \ion{H}{I} morphology towards parts of the IV Arch. \cite{panopoulou2019a} perform a tomographic decomposition of stellar polarization as a function of distance towards a sightline at intermediate galactic latitude that intercepts LVC and IVC gas. They find that the inferred plane-of-sky magnetic field orientation differs by 60$^\circ$ between the LVC and the IVC. Additional evidence comes from the Faraday rotation map of \cite{opperman}. The IV Arch is found to correlate with a large region of positive Faraday depth, which could point to a systematic difference in the line-of-sight component of the magnetic field compared to local gas. \citet{tritsis2019} apply a novel method to estimate the plane-of-sky magnetic field strength in LVCs and IVCs towards Ursa Major. They find magnetic field strengths that vary both on the plane of the sky and along the line of sight. These results are not surprising, as the distance to most IVCs (of order 1 kpc) is much larger than the correlation length of the Galactic magnetic field \citep[$\sim 200$ pc, e.g.][]{beck2016}.

In Section \ref{ssec:ivc-correlations} we find tentative evidence that IVCs may be contributing to the polarized dust emission at 353 GHz (by adding a depolarizing effect). This is in apparent contrast with the finding of \cite{skalidis2019} that at high latitudes the \textit{Planck} polarized intensity is dominated by dust in the Local Bubble wall. These authors compared the \textit{Planck} polarized dust emission with starlight polarization at different distances. They found that the match between the two tracers at high galactic latitude is best at distances of 200-300 pc. However, as noted by the authors, the stellar polarization sample does not uniformly cover the high latitude sky. In particular there is lack of stellar measurements at distances further than 400 pc towards the general area occupied by IVCs \cite[see Figure D.1 in ][]{skalidis2019}. Future stellar polarization surveys at high latitude \citep[e.g.][]{pasiphae} will help clarify what fraction of the polarized dust emission arises from further than the Local Bubble wall.

The large-angular-scale coverage of IVCs, combined with their common astrophysical origin \citep[likely a Galactic fountain process][]{kunzdanly} may act to produce a systematic effect in the polarized dust SED (in the form of frequency decorrelation). So far, frequency decorrelation remains undetected in the \textit{Planck} data \citep{planckdecorr, sheehy}. However, if such a signal is to be searched for in the high-Galactic latitude sky, the regions with significant IVC contribution to the column density would present a prime target. The maps of IVC column density presented in this work could serve as templates for forward modeling the polarized dust emission SED in order to predict the level of decorrelation that is to be expected due to their presence. Another possible avenue of investigation would be to perform parametric fits to the polarization data that explicitly take into account contributions from IVCs and LVCs separately, as has been done for total intensity \citep{planck2011_xxiv}.

As the required accuracy of polarized foreground modeling increases, analyses that take into account IVCs may also serve for testing/improving modeling of Galactic synchrotron emission (the dominant low-frequency CMB foreground). Current models assume a single correlation parameter between the synchrotron and dust emission polarized signals \citep{BICEP2018,planckdecorr}. However, this assumption would not be accurate if the detected synchrotron emission and dust emission did not probe the same path-length over all sightlines. 
For example, while dust emission from IVCs might dominate that of local gas in some sightlines, synchrotron emission from the Galactic halo could be suppressed compared to local emission. It is known that the cosmic ray density is significantly lower towards certain IVCs compared to the Galactic plane \citep{tibaldo2015}. The magnetic field in the Galactic halo is also observed to be lower than in the Galactic plane \citep[specifically, the line-of-sight component of the field][]{sobey2019}. Determining whether the polarized synchrotron emission at the location of IVCs is detectable (and if so, at what frequency) would help inform choices in combined dust-synchrotron foreground models.

\subsection{Limitations of the current work}
\label{ssec:limitations}

\subsubsection{Spatial coherence}
Treating HEALPix superpixels independently means there is no imposed spatial coherence at scales larger than the superpixel size. This can lead to abrupt changes between pixels. Discontinuities can arise when for example two components that are nearby in velocity appear distinct in one pixel but are blended in a neighboring pixel. The method has no way of distinguishing the two in the latter case. The fact that our maps do show large-scale continuity in the number of identified clouds per pixel, indicates that there exist multiple distinct velocity components that are separate enough in velocity for the algorithm to identify them individually. 

In Fig. \ref{fig:chi2} the IVC column density map shows discontinuities towards certain pixels. These discontinuities arise from the selection of the LVC-IVC velocity boundary. It happens that in these regions a cloud is peaked very near the boundary of -12 km/s. In some pixels the velocity centroid of the cloud falls in the IVC range and in neighboring pixels it is found in the LVC range. An improved IVC column density map could be constructed by iteratively finding the IVC/LVC boundary by considering not only the global statistics of clouds (Fig. \ref{fig:nhvel}) but also the local properties of clouds in a region. Methods that take into account solutions in neighboring pixels could be adjusted for this \citep[e.g.][]{green2019}.

\subsubsection{Confusion in velocity}

\begin{figure}
\includegraphics[scale=1]{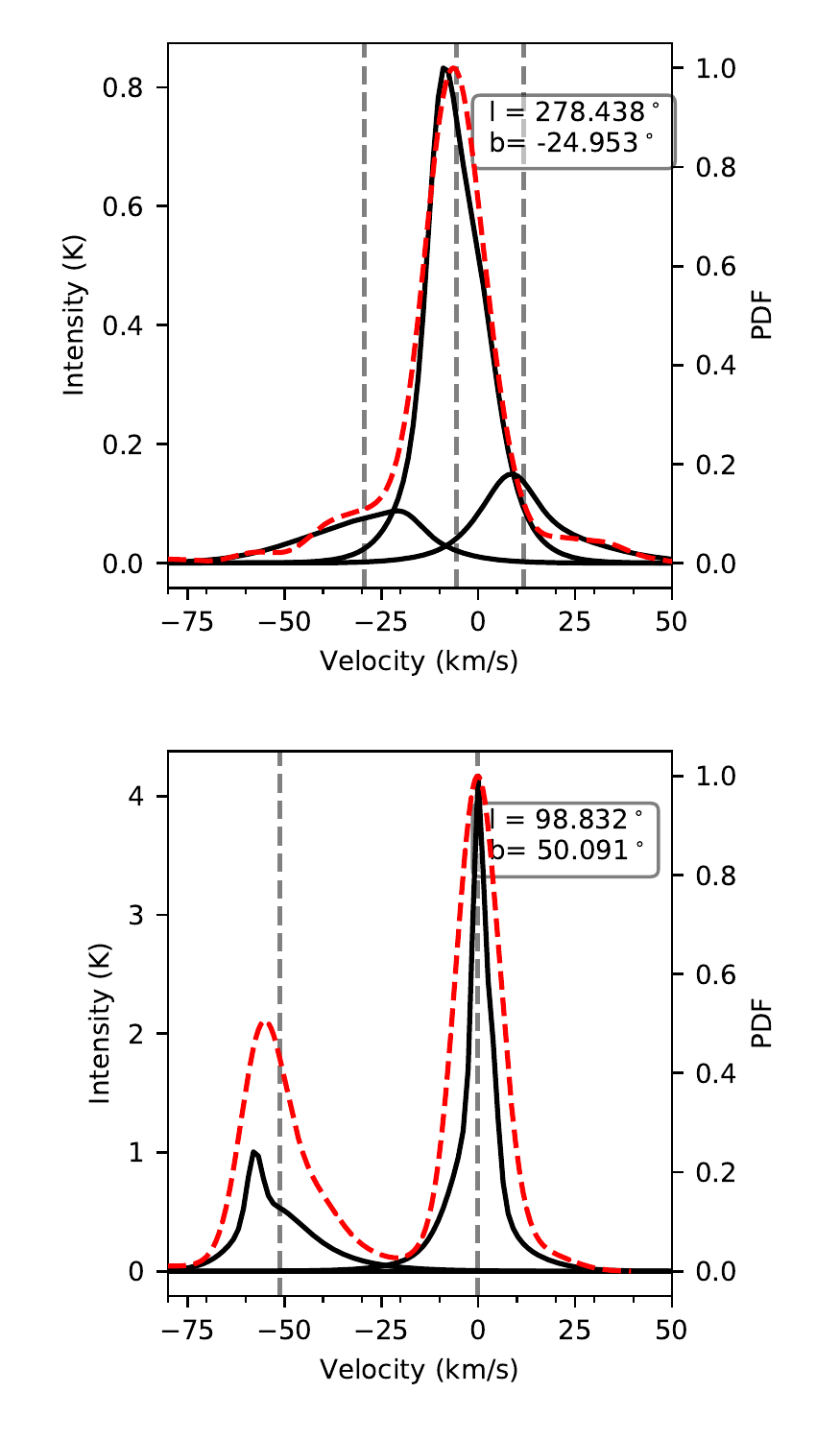}
\caption{Top: Example of mean spectrum of a cloud that suffers from confusion with signal from a nearby bright peak (cloud spectrum, left most black line) Bottom: Example of mean cloud spectrum with substructure that has not been detected by the method (second from the right black line). The dashed red line shows the PDF of Gaussian Component velocities in the specific superpixel. Vertical gray lines mark the centroid velocity of each cloud. The boxes on the top right corner show the coordinates of the superpixel.}
\label{fig:skew}
\end{figure}

As explained in Section \ref{sec:method}, we smooth the spectral information of the original data during the construction of the PDF of Gaussian component velocities with a KDE. This means that we are losing the ability to identify clouds that are separated in velocity by less than $2.5\times 4$ channels ($\sim 12$ km/s). Pixels with distinct peaks that differ by less than the bandwidth will be identified by the method as a single cloud. An example is shown in Figure \ref{fig:skew} (bottom panel). A single peak in the PDF is made of Gaussians with a bimodal distribution of velocity, leading to a double-peaked spectrum for the 'cloud'. This effect can be mitigated by using a smaller KDE bandwidth, at the expense of introducing more spurious cloud identifications (Section \ref{sec:validation}). 

A small fraction of the fainter clouds identified by the method are subject to certain systematics. First, in the North, there exists emission from residual stray radiation. Even though this issue is addressed by our pre-processing routine (Appendix \ref{sec:appendixA}), it is likely that some part of this spurious signal remains and is attributed to clouds. 
Second, a problem occurs when a strong peak in the PDF is located near a fainter peak (e.g. Figure \ref{fig:skew}, top). Gaussians that belong to the stronger peak may lie at a velocity that has been assigned to the weaker peak. As a result, the fainter cloud may show a skewed mean spectrum. This affects mostly faint clouds that are adjacent to very bright ones in velocity space. This does not affect our estimates of the number of clouds per pixel. Its main effect is on the estimate of $N_{\ion{H}{I}}$ of fainter clouds, biasing it to slightly higher values.

We quantify the percentage of clouds that are affected by this latter issue by calculating a measure of skewness of the spectrum of each cloud. We calculate the difference $\delta$ between the centroid velocity and the peak velocity of the mean spectrum of each cloud. The values of $\delta$ are in the range [-12,12] channels, with the 10- and 90- percentiles being -2 and -1.9 channels. The most extreme values of $\delta$ ($|\delta| > 5$ channels) are found in 80\% of cases in clouds with $N_{\ion{H}{I}} < 8 \times 10^{19} {\rm cm}^{-2}$. Clouds of even lower column density $N_{\ion{H}{I}} < 1 \times 10^{19} {\rm cm}^{-2}$ make up 65\% of extreme-$\delta$ cases.

Our method can identify components of emission that are kinematicaly distinct. It is worth noting that the velocity differences found in this work are much larger (typically tens of km/s) than what is expected to arise from natural velocity dispersion (thermal/turbulence) within individual clouds. For gas temperatures of 100 - 1000 K, the isothermal sound speed is 1 - 3 km/s. While different velocity components do not necessarily map to discrete structures along the line-of-sight \citep[][]{Beaumont,clarkeppv}, any velocity substructure is smoothed out due to (a) the spectral resolution of the HI4PI data ($\sim$ 1 km/s) and (b) our KDE smoothing kernel (5 km/s). 

It is possible that multiple clouds exist within one pixel that cannot be separated in velocity. Because of this, our estimate of the number of clouds per pixel is a \textit{lower limit}. This limitation can be surpassed by using complementary methods and data, for example by mapping stellar reddening as a function of distance (see references in the Introduction), or by searching for \ion{H}{i} filaments with different orientations in different velocities \citep{clark2018}. The present method can be used in conjunction with such approaches in order to improve reconstruction of the 3D ISM.

\subsection{Future directions}

The data presented here can be used to improve 3D maps of reddening at high Galactic latitude. For example, the number of clouds per pixel can be used as a prior in fitting the line-of-sight reddening profile in models like that of \cite{Zuckerdistances}. Additionally, the IVC and LVC column density maps can be used as spatial templates, extending the analysis used for molecular cloud data to the diffuse ISM \citep[][]{zucker}. Knowledge of the number of components along the line of sight and their properties in combination with starlight polarization can also facilitate 3D mapping of the ISM magnetic field \citep{panopoulou2019a}.

Given the increasing sensitivity of ongoing and upcoming CMB experiments, there is an urgent need for modeling dust foregrounds with unprecedented precision \citep[at the nano-Kelvin level][]{litebird,CMBS4}. Our results can be used to inform more realistic models of polarized dust emission in the near future. For example, the number of clouds per line of sight can serve as an informative prior for parametric component separation methods \citep[e.g.][]{commander}, where the foreground signal is modeled independently in each pixel of the sky.

\section{Data products}
\label{sec:product}

We provide data\footnote{\url{https://doi.org/10.7910/DVN/8DA5LH}} in the following forms:
\begin{itemize}
    \item We provide two HEALPix FITS files for the polar areas studied in Section \ref{ssec:polarcaps} and two HEALPix FITS files for the BICEP/Keck region (Section \ref{sec:bicep}). For each region, one file contains the column density maps: $N\rm _{\ion{H}{i}}$, $N \rm ^{LVC}_{\ion{H}{i}}$, $N\rm ^{IVC}_{\ion{H}{i}}$. The other file contains maps of the number of cloud statistics: $\rm N_{clouds}$ and $\mathcal{N}_c$. The maps are given at a resolution of $N_{\rm{side}} =$ 128, corresponding to a pixel size of 0.46$^\circ$ on each side.
    \item For each of the studied regions, we provide a file in hdf5 format that holds properties of the entire sample of clouds (including velocities that were excluded for the analysis in Section \ref{sec:results}). The reported properties are: the cloud's column density ($N\rm ^{Cloud}_\ion{H}{i}$), the superpixel index in which it belongs, the centroid velocity ($\rm v^{Cloud}$) and second moment ($ \delta {\rm v^{Cloud}}$) of the cloud spectrum, the number of Gaussians that make up the cloud, and the number of maxima in the cloud spectrum).
\end{itemize}

The data are provided for the optimal choice of bandwidth parameter (4 channels, or 5 km/s). A Python implementation of the method discussed in Section \ref{sec:method} is publicly available on github\footnote{\url{https://github.com/ginleaf/cloudcount}}. The data are accompanied by a Python a tutorial for using the data products.

We caution against smoothing the $\rm N_{clouds}$ and $\mathcal{N}_c$ maps to obtain lower resolution versions. The reason is simply illustrated with the following example. Let us suppose that one pixel is occupied by a single cloud ($c_1$), while its neighboring pixel contains 2 clouds ($c'_1,c'_2$). The value for the number of clouds in the area that covers both pixels would be 1.5 if we naively averaged $\rm N_{clouds}$ in the region, which is clearly in error. An alternative would be to assign the value of $\rm N_{clouds}$ that is maximum among the superpixels that make up the area (giving 2 clouds in this case). This would give the correct result if $c_1$ is the same cloud as either $c'_1$, or $c'_2$, but an incorrect result if the three clouds are all distinct from each other. Lower resolution products are created by repeating the analysis of Section \ref{sec:method} with a different choice of  $N_{\rm{side}}$ and can be made available upon request.

\section{Summary}
\label{sec:conclusion}

We have developed a method to identify clouds of \ion{H}{i}, defined as kinematicaly distinct components of emission. The method uses a Gaussian decomposition of \ion{H}{i} spectra and searches for overdensities in the PDF of Gaussian component velocities within HEALPix superpixels of size $\sim$ 0.5$^\circ$ ($N_{\rm{side}}$ 128). The value of the main parameter of the method, the kernel bandwidth, is optimized through tests with real and mock data. These tests show that the method identifies clouds at the correct velocity for over 85\% of cases with a false positive rate of less than 10\% (Appendix \ref{sec:validation}).

We implemented this method for the high Galactic latitude sky. The method does well in assigning the majority of the emission to clouds, with very little residual signal left out (median relative residuals of 0.5\%, Appendix \ref{sec:validation}). We present maps of the number of components along the line of sight and statistical properties of the identified clouds. We analyse clouds in the velocity range $\rm |v_{LSR}| <$70 km/s. Throughout the high-latitude sky, the number of clouds per pixel is small: less than 6 clouds per pixel are found to contribute to the \ion{H}{I} column. The Northern sky has on average a larger number of clouds per pixel (3) than the Southern (2.5). This is mostly due to the prominence of IVCs in the North. 

We introduced a measure of the number of clouds that takes into account the column density of different clouds along the line of sight. This quantity, $\mathcal{N}_c$, is more robust to uncertainties in the method. We find that the majority of pixels at high latitude have a low value of $\mathcal{N}_c < 1.5$ (60\% in the North and 90\% in the South). The statistics of the number of clouds per pixel are dominated by IVCs, which show large-scale spatial coherence, especially over the Northern sky. We find that the presence of IVCs affects the fit to the dust emission SED by \textit{Planck}. 

We also implemented the method on the region targeted by the BICEP/Keck CMB experiment. We find evidence for multiple components along the line of sight. However, for most of the area the column density is dominated by a single component in each pixel. We find that regions with elevated $\chi^2$ from the fit to the dust SED show evidence for more complex kinematics in the \ion{H}{i} gas.

We discuss our results in the context of CMB foreground modeling and highlight the potential importance of IVCs. Our results can aid in informing future 3D dust mapping efforts as well as CMB foreground analyses. The data are made publicly available as described in Section \ref{sec:product}.

\acknowledgments

The authors thank Kostas Tassis and Vincent Pelgrims for comments on the paper. G. V. P. thanks Alberto Krone-Martins, Marc-Antoine Miville-Desch{\^e}nes and Brandon Hensley for helpful discussions. We thank Vincent Guillet for suggesting the use of a column-density-weighted number of clouds metric. We thank Mathieu Remazeilles for providing the $\chi^2$ map of the dust model from Planck. G. V. P. acknowledges support from the National Science Foundation, under grant number AST-1611547. Support for this work was provided by NASA through the NASA Hubble
Fellowship grant \# HST-HF2-51444.001-A awarded by the Space Telescope
Science Institute, which is operated by the Association of Universities for
Research in Astronomy, Incorporated, under NASA contract NAS5-
26555. This work made use of healpy \citep{healpy} and Astropy \citep{astropy:2013,astropy:2018}.

\vspace{5mm}
\facilities{Effelsberg, Parkes, Planck}


\software{astropy \citep{astropy:2013},  
          healpy \citep{healpy}, 
          }


\appendix

\pagebreak

\section{Removing contamination from uncorrected stray radiation}
\label{sec:appendixA}

\begin{figure*}
\centering
\includegraphics[scale=0.6]{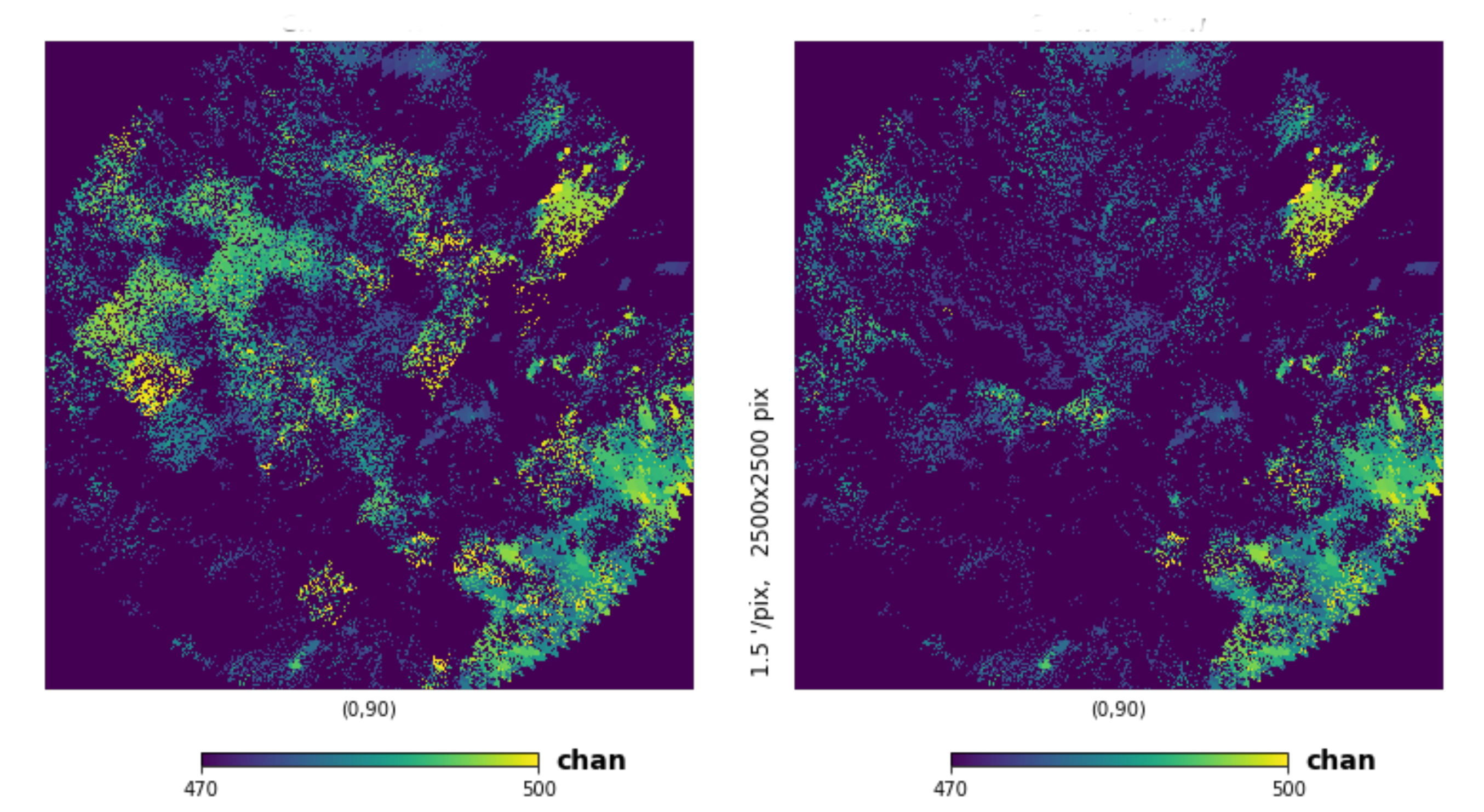}
\caption{\textit{Effect of removing stray radiation patterns.} Map of cloud velocity in North Galactic Polar cap (b $> $ 60 deg) without (left) and with (right) removal of patterns (only showing velocity channels where the patterns appear).}
\label{fig:squares}
\end{figure*}

The part of the HI4PI survey that was conducted from the Northern hemisphere (Effelsberg-Bonn HI Survey, EBHIS) contains low-amplitude artifacts that originate from residual uncorrected stray radiation. These artifacts appear as faint increases in brightness with a distinctive square pattern (approximately 5$^\circ \times 5 ^\circ$, reflecting the scanning pattern). They are usually visible at velocities where Galactic emission is very faint or absent. See also \cite{ghigls} for more details. 

These artifacts have not been removed prior to performing the Gaussian decomposition in Lenz et al. (in prep). Gaussians associated with these artifacts will result in detections of spurious `clouds' by our method. To avoid this, we pre-process the Gaussian decomposition in order to remove as much contamination as possible. 

The artifacts were identified by eye in the spectral channel maps of the EBHIS part of the survey. We removed Gaussian components that were associated with artifacts by the following process:
\begin{itemize}
    \item[1.] For each noise square, we create a list of all pixels within a circular region of 6 degrees diameter.
    \item[2.] We flag any Gaussian component in the decomposition if (a) it is within a region affected by noise (defined as above) AND (b) if its mean is within 0.5 sigma of the velocity channel where a noise square was identified.
    \item[3.] All flagged Gaussian components are removed before proceeding with the analysis. 
\end{itemize} 

Figure \ref{fig:squares} compares the results of the cloud identification method before and after performing the pre-processing step. We compare the map of the mean velocity of clouds identified in a region centered on the North Galactic Pole (with a radius of 30 degrees). The map shows only clouds within a range of velocities where artifacts are prominent (40-50 km/s). The run with pre-processing (right panel) effectively eliminates the square-patterned residuals present when no pre-processing is applied (left panel). We are confident that our method removes all artifacts with maximum amplitude more than 0.19 K and rms (within the square) of 0.12 K or higher.

\section{Validation tests \& parameter optimization}
\label{sec:validation}

\subsection{The KDE bandwidth parameter}

\begin{figure}
\centering
\includegraphics[scale=1]{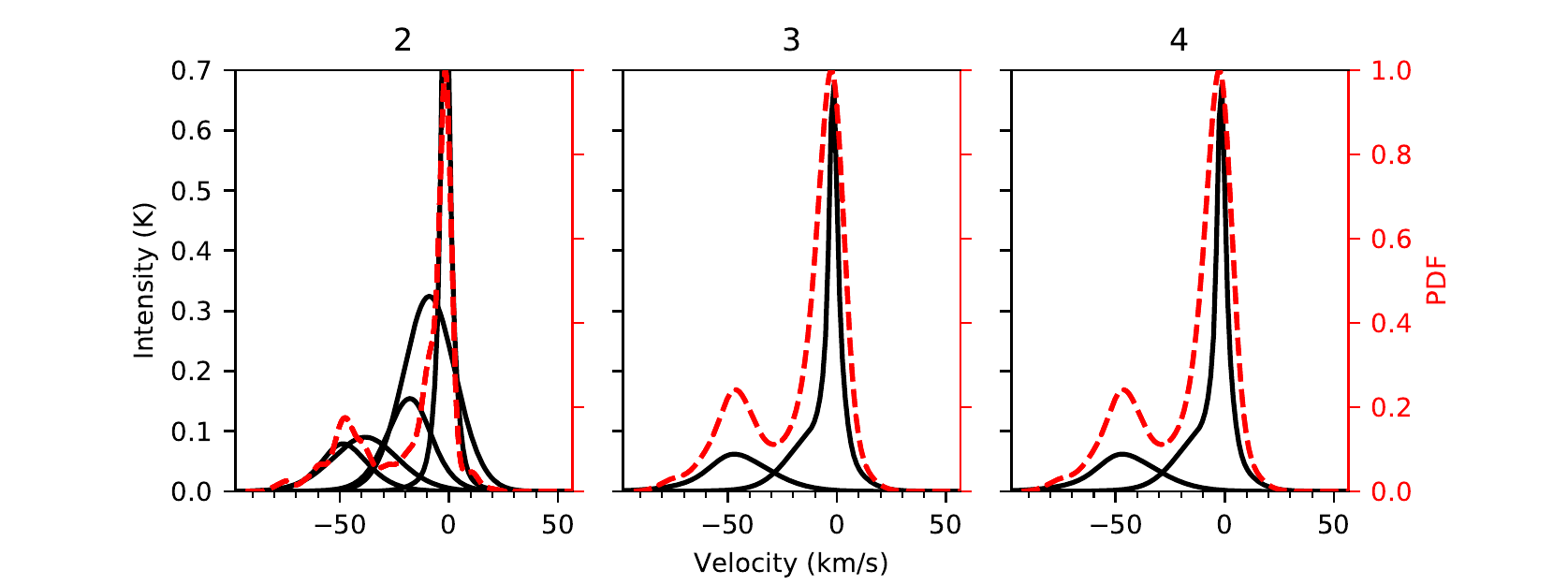}
\caption{\textit{Effect of KDE bandwidth on clouds found in real data.} From left to right: bandwidth of 2, 3 and 4 channels. The red dashed line is the resulting PDF of the mean velocity of Gaussians in one $N_{\rm{side}}$ 128 pixel. A black line shows the sum of the emission from all Gaussians that have been identified as belonging to the same cloud (there are more than one clouds in this sightline). The selected pixel is centered on l,b = (56.09$^\circ$, 63.07$^\circ$). }
\label{fig:bandwidth-spec}
\end{figure}

The method presented in Section \ref{sec:method} constructs the PDF of Gaussian component velocities by use of a KDE. The choice of KDE size (bandwidth) determines the resolution of the cloud identification in velocity. Figure \ref{fig:bandwidth-spec} shows the effect of varying the value of this parameter (from 2 to 4 channels) on the resulting clouds. For the finest resolution, there are spurious features in the PDF. Since the method locates clouds as separate peaks in the PDF, the spurious local maxima found when using a small bandwidth lead to the 'detection' of many peaks. However, the slightly larger bandwidths of 3 and 4 channels produce smoother PDFs that lack these spurious detections. The cloud identification for the bandwidth of 3 channels is in agreement with that of 4 channels for this selected pixel.

We investigate the optimal choice of KDE bandwidth through the following test. We generate mock data of Gaussian velocities for $10^4$ superpixels. Each 1D distribution of velocities has properties similar to what can be found in a superpixel of the HI4PI dataset at high Galactic latitude. For each superpixel, we first determine the number of peaks in the distribution of Gaussian velocities (clouds) by drawing from a skewed-normal distribution that peaks at 2 components, with a tail out to 8 components. The number of Gaussians that belong to each cloud is drawn from a normal distribution centered on 10, with a standard deviation of 400. We take the absolute value of the distribution to ensure positive-definite numbers. We assume that each cloud has a distribution of Gaussian Component (GC) velocities that is normal. The mean velocity for each cloud is drawn from a normal centered on -30 km/s, with a standard deviation of 40 km/s, covering the entire range of velocities observed in the high latitude sky, with the exception of HVCs. The standard deviation is drawn from a uniform distribution in the range [4,13] km/s. We repeat this process 10$^4$ times, thus generating mock data for 10$^4$ superpixels. 

\begin{figure*}
\centering
\includegraphics[scale=1]{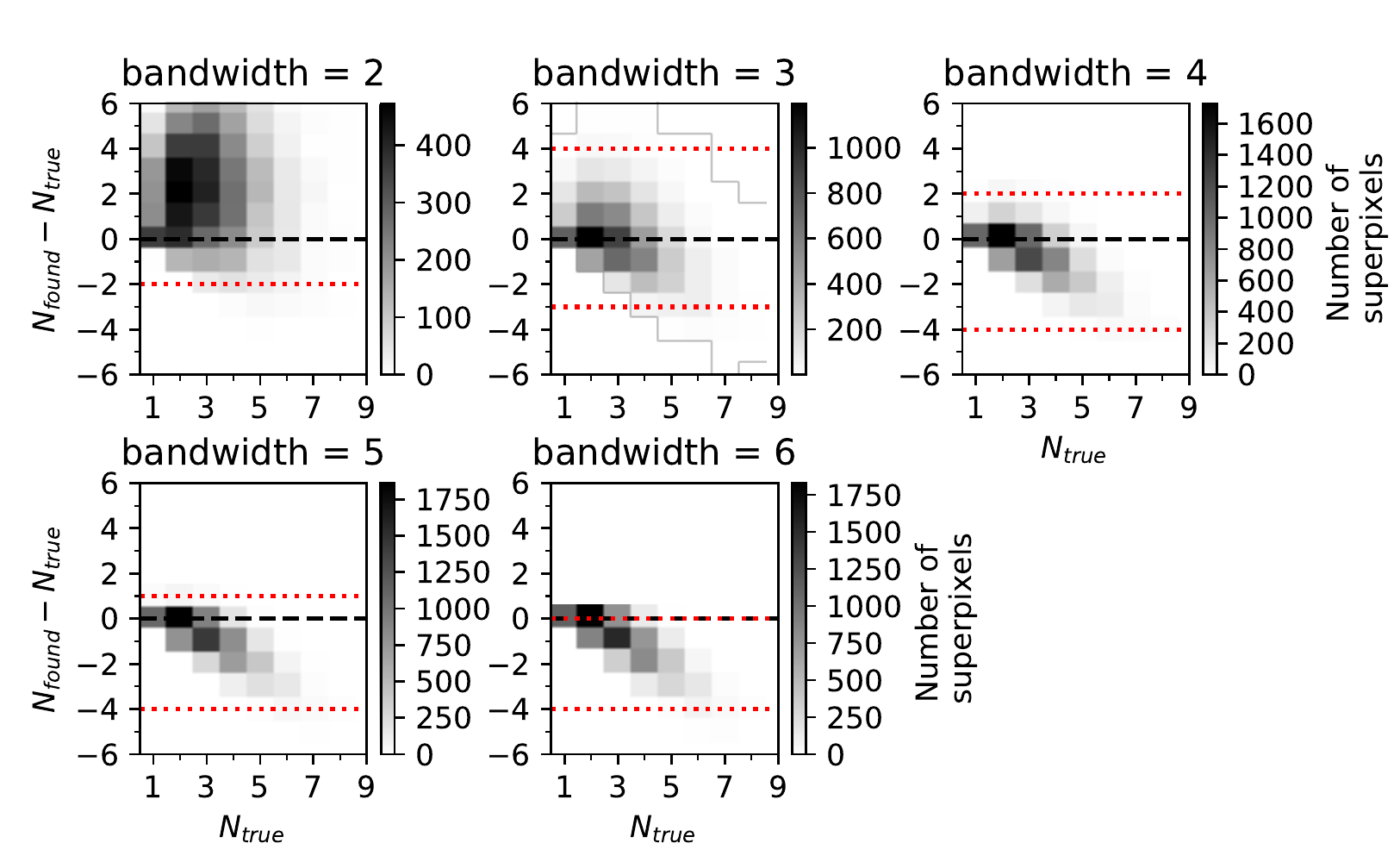}
\caption{\textit{Evaluation of the KDE bandwidth from tests on mock distributions of Gaussian parameters.} The difference between the number of peaks found by the method, $N_{found}$, and the true number of peaks ($N_{true}$) as function of $N_{true}$. Each panel shows the results for a different value of the KDE bandwidth (2, 3, 4, 5 and 6 channels). A horizontal black dashed line marks 0 (perfect match with true solution).The horizontal red lines mark the 1$-$ and 99$-$ percentile of the distribution of $N_{found}-N_{true}$ (for bandwidth = 2 the 99$-$percentile is outside the shown range). Larger bandwidths result in less spurious detections of clouds at the cost of reduced spectral resolution.}
\label{fig:validation_bandwidth}
\end{figure*}

We then run the cloud identification code with different values of KDE bandwidth (from 2 channels to 6 channels in steps of 1). One way of evaluating the method's success is to compare the number of peaks it has identified in a single pixel with the true number of peaks (input in generating the mock data for that pixel). Figure \ref{fig:validation_bandwidth} shows the difference between the recovered number of peaks ($N_{found}$) and the true number of peaks ($N_{true}$) as a function of the latter, for all pixels in the test. When a bandwidth of 2 is used, the code finds more peaks than were originally input. These are spurious peaks due to the noisiness of the PDF. As the bandwidth increases, less and less of such spurious peaks are found. At the same time, however, the reduced spectral resolution when using larger bandwidths means that some peaks that were nearby in velocity space cannot be separated, and are counted as a single peak. This is why at a bandwidth of 6, there are many pixels in which the method finds less peaks than were input. There is a trade-off in the choice of bandwidth. We opt for an intermediate bandwidth: one that minimizes the amount of spurious peaks, while at the same time not significantly compromising the spectral resolution.

A second way of evaluating the method's success is by measuring how well the peak velocities of recovered clouds compare to the input peak velocities. For this we calculate the difference between the mean velocity of each recovered cloud with its nearest (in velocity) counterpart in the original input of the test. We find that there is no bias, the distribution of velocity differences has a mean of 0 channels for all bandwidths used. However, the distribution shows wide wings for a bandwidth of 2 channels, something that is reduced for larger bandwidths. The 38-percentile of the distribution of velocity differences is 4 channels for a bandwidth of 2. For larger bandwidths (3,4,5,6) 4 channels corresponds to the 64, 85, 93, 96 percentile. Thus, by choosing a bandwidth of 4 or larger, more than 85\% of clouds will be identified by the method within 4 channels of their true value. The percentage of clouds that will be identified within 2 channels of their true value is 75\% for the same choice of bandwidth.

We also examine the false positive rate, that is the fraction of peaks that have no true peak within their assigned range of velocities. This fraction is 70\%, 30\%, 10\% at bandwidths of 2, 3, 4 channels and drops to 3\%, 1\% at bandwidths of 5 and 6 channels, respectively.

From the above tests, we conclude that a bandwidth of 4 is optimal: it combines high enough spectral resolution for identifying separate peaks while minimizing spurious detections and ensures that the recovered clouds will be located within a few channels of their true value.

\subsection{The $N_{\rm{side}}$ parameter}

The maps presented in the main text were created by applying the method to superpixels of $N_{\rm{side}}$ = 128. Here we repeat the analysis of the high-galactic latitude sky to investigate the effect of changing the $N_{\rm{side}}$ to 64 (the immediately lower resolution in a HEALPix pixelization).

We expect our method to work well as long as there is a statistically large enough sample of Gaussians per pixel that are used to create the velocity PDF. This is true for both resolution choices. For $N_{\rm{side}}$ = 64, the number of Gaussians per pixel ranges from 300 to 2480 for the Southern Galactic cap. In 80\% of pixels more than 880 Gaussians are used to construct the PDF.
For the higher resolution of $N_{\rm{side}}$ = 128, the minimum number of Gaussians per pixel is 68 and the maximum is 664. 

\begin{figure*}
\centering
\includegraphics[scale=1]{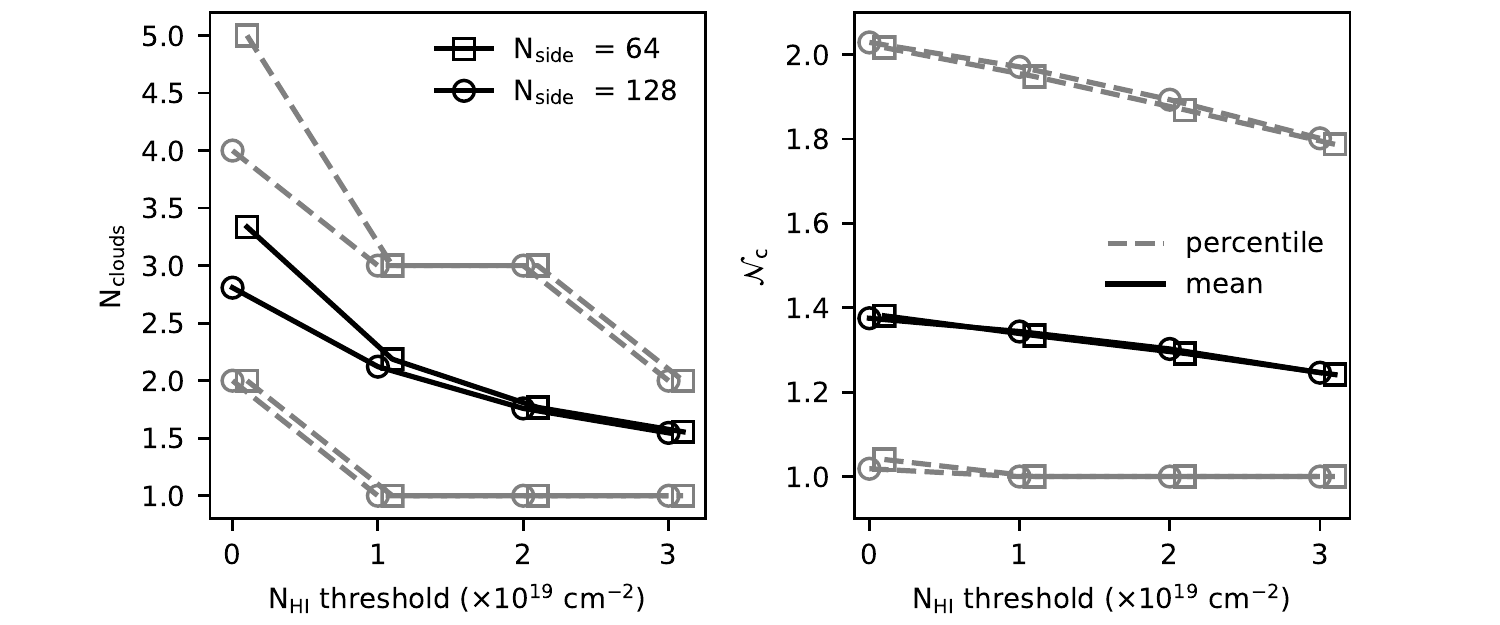}
\caption{\textit{Effect of the choice of $ N_{\rm side}$ on the distribution of $\rm N_{clouds}$ and $\mathcal{N}_c$ for the high-galactic latitude sky.} Left: The mean (black line), 10- and 90- percentile (gray dashed lines) of the distribution of $\rm N_{clouds}$ for different values of threshold on cloud $\rm N_{\ion{H}{i}}$. Results for $N_{\rm side} = 128$ are shown with circles while those for $N_{\rm side} = 64$ are shown with squares. The points for $N_{\rm side} = 64$ are displayed to the right of those for $N_{\rm side} = 128$ for better visualization. The results for different $N_{\rm side}$ agree when discarding low-$\rm N_{\ion{H}{i}}$ clouds. Right: As in the left panel but for the distribution of $\mathcal{N}_c$. The distribution of $\mathcal{N}_c$ is largely insensitive to the choice of resolution.}
\label{fig:converge}
\end{figure*}

First we compare the distributions of $\rm N_{clouds}$ that result from applying the method with a superpixel $N_{\rm side}$ of 64 and 128. The mean, 10-percentile and 90-percentile are 2.8, 2 and 4, respectively, for $N_{\rm side} = 128$. These values change to 3.3, 2 and 5 for $N_{\rm side} = 64$. The main reason for these differences is that small changes in the PDF (that can arise from different superpixel size choices) can easily affect the detection (or not) of low $N_{\ion{H}{I}}$ clouds. By imposing a threshold on the cloud $\rm N_{\ion{H}{i}}$ we find improved agreement between the distribution of $\rm N_{clouds}$ at different $N_{\rm side}$. This is shown in Figure \ref{fig:converge} (left).

Next we compare the distribution of $\mathcal{N}_c$ for the two resolutions (Figure \ref{fig:converge}, right). The distributions at different resolution agree for all values of the $\rm N_{\ion{H}{i}}$ threshold. This is to be expected, as $\mathcal{N}_c$ down-weighs low-column-density clouds, which are sensitive to the choice of superpixel size. Visual inspection of the $\mathcal{N}_c$ maps shows the same large scale features arising for both choices of superpixel size. 

\begin{figure*}
\centering
\includegraphics[scale=1]{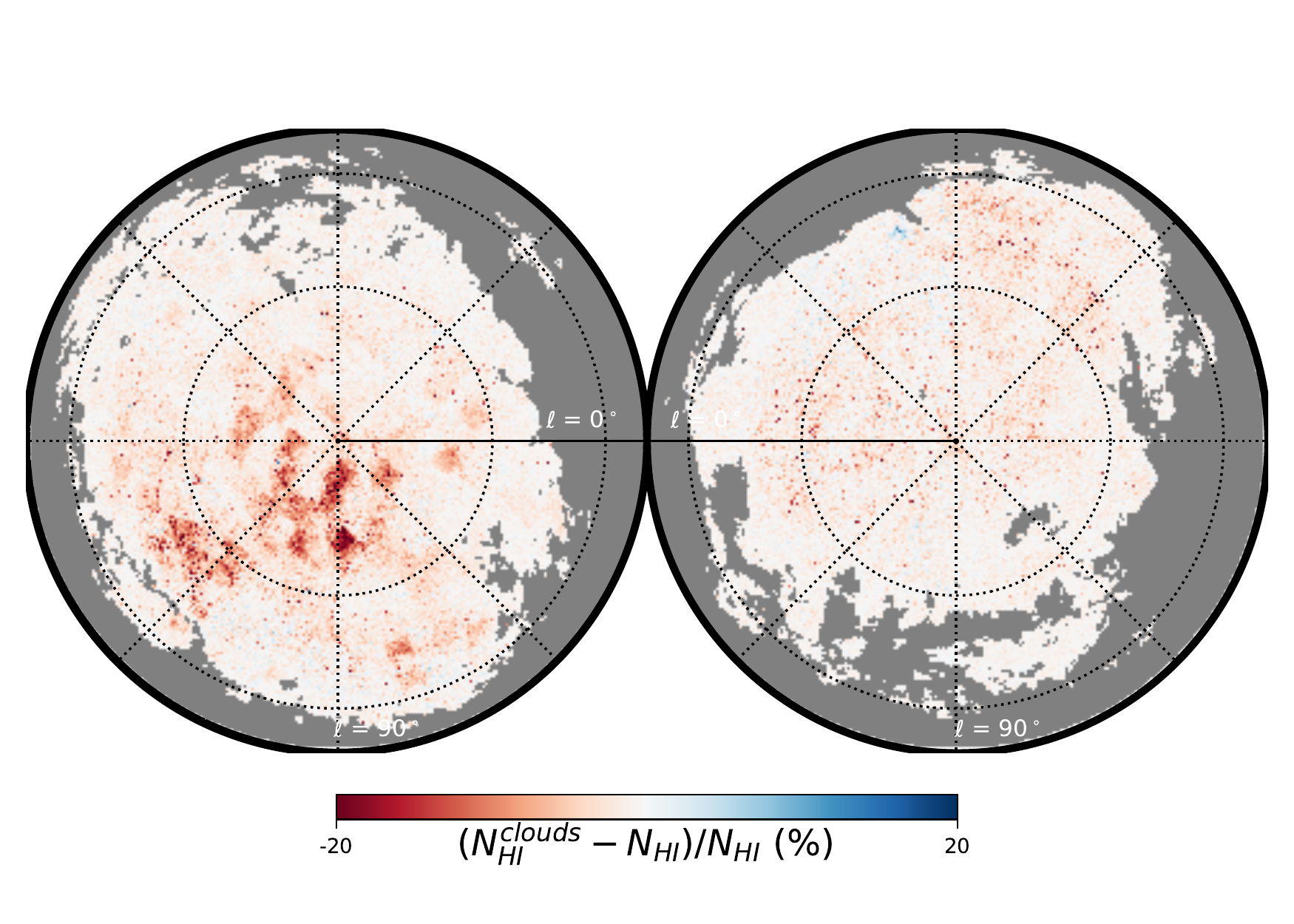}
\caption{Relative residual NH map, showing the excess at the locations of the removed noise squares (projection as in Fig. \ref{fig:nclouds_map}).}
\label{fig:residmap}
\end{figure*}

\subsection{Evaluation of residuals}

The final step in validating our method is to examine if there is emission that is not picked up by our cloud identification. Figure \ref{fig:residmap} shows the difference between the integrated emission of the HI4PI data and the integrated emission from the Gaussian components assigned to clouds for pixels in the North and South sky regions.
We find very small residuals, with a median of -0.8\%. Only 1.2\% of pixels have a relative residual $\Delta_{NH,rel} < -10\%$ and 0.005\% have $\Delta_{NH,rel} > 10\%$.
This shows that the clouds we identified are responsible for all but a negligible fraction of the total extinction. In the residual map, the regions where subtraction of stray light radiation residuals (Appendix \ref{sec:appendixA}) appear as patches of underestimated column density (by 10-20\%). Regions where cloud $N_{\ion{H}{I}}$ overestimates the total column by more than 20\% are associated with point sources or the Magellanic system, where there is significant signal outside the range of velocities used in the Gaussian decomposition. Additionally, the $N_{\ion{H}{I}}$ calculated by integrating the \ion{H}{I} spectra may in rare cases include summation over negative values, due to presumably improper baseline subtraction. The GCs are forced to have positive amplitude and this causes by default overestimation. This effect is minor.

\bibliography{draft}{}
\bibliographystyle{aasjournal}
\end{document}